\documentstyle[epsfig]{mn2e}
\begin{document}

\newcommand{\science}{{ Science}}
\newcommand{\physrev}{{ Phys. Rev.}}
\newcommand{\pasp}{{ PASP}}
\newcommand{\prd}{{ Phys. Rev. D}}
\newcommand{\apj}{{ ApJ}}
\newcommand{\aj}{{ AJ}}
\newcommand{\apjl}{{ ApJL}}
\newcommand{\apjs}{{ApJS}}
\newcommand{\aplc}{{\em Astrophys. Lett. Comm.}}
\newcommand{\prl}{{PRL}}
\newcommand{\mnras}{{MNRAS}}
\newcommand{\araa}{{ ARA\&A }}
\newcommand{\aap}{{ A\&A}}
\newcommand{\aaps}{{ A\&A Supp}}
\newcommand{\pasj}{{ PASJ}}
\newcommand{\nat}{{ Nature}}
\newcommand{\arcseca}{\,\rm{arcsec}}

\title [Discs in early-type lens galaxies] {Discs in early-type lensing
galaxies: effects on magnification ratios and measurements of $H_0$} \author [Ole M\"oller, P. Hewett \& A.W. Blain] { Ole
M\"oller$^1$, Paul Hewett$^2$ and A. W. Blain$^3$\\ $^1$Kapteyn
Institute, PO Box 800, 9700 AV Groningen, the Netherlands.\\
$^2$Institute of Astronomy, Madingley Road, Cambridge CB3 0HE, UK.\\
$^3$Department of Astronomy, California Institute of Technology,
CA91125, Pasadena, USA\\ } \maketitle


\begin{abstract} Observations of early-type galaxies, both in the local universe
and in clusters at medium redshifts, suggest that these galaxies often contain
discs or disc-like structures.  Using the results of Kelson et al.  (2000)
for the incidence of disc-components among the galaxies in
the redshift $z=0.33$ cluster CL 1358+62, we investigate the effect of disc
structures on the lensing properties of early-type galaxies.  Statistical
properties, like magnification cross sections and the expected number of quad
(four-image) lens systems, are not affected greatly by the inclusion of discs
that contain less than $\sim$ 10\,per cent of the total stellar mass.  However,
the properties of individual lens systems are affected.  We estimate that
$10-30$\,per cent of all quad lens systems, with early-type deflector galaxies,
would be affected measurably by the presence of disc components.  Intriguingly,
the image magnification ratios are altered significantly.  The amplitude of the
predicted change is sufficient to explain the observed magnification ratios in
systems like B1422+231 without requiring compact substructure.  Furthermore,
time delays between images also change; fitting a bulge-only model to early-type
lenses that in fact contain a disc would yield a value of the Hubble constant
$H_0$ that is systematically too low by $\sim25$\,per cent.

\end{abstract}


\section{Introduction}

In current models of hierarchical galaxy formation the morphological type of a
galaxy is determined to a large extent by its merger history
\nocite{abraham2001}({Abraham} \& {van den Bergh} 2001):  late-type galaxies
have not undergone a recent massive merger, so that stable, prominent discs
formed, whereas early-type galaxies have been prevented from forming a prominent
disc by a major merger that either consumes all the available gas for star
formation, or redistributes angular momentum via large scale winds
\nocite{somerville1999}({Somerville} \& {Primack} 1999).  This behaviour has
also been recovered in hydrodynamical simulations of galaxy formation
\nocite{steinmetz1995}({Steinmetz} \& {M\"uller} 1995).  So far, there is no
evidence for any evolution in the luminosity and size of discs in late-type
galaxies up to redshifts of about 1 \nocite{simard1999,ferguson2001}({Simard}
{et~al.}  1999; {Ferguson} \& {Clarke} 2001).  This suggests that if mergers
occur after that time, they are either not massive enough to significantly
affect the disc structures or they destroy visible disc structures, producing
elliptical systems.  It is not known on what time scales disc structures form.
Simulations and some observations suggest that disc structures form slowly between redshifts 0.1 and 1
\nocite{mo1998}({Mo}, {Mao} \& {White} 1998)  but there is also
evidence that some large discs are already present at higher redshifts
\nocite{dokkum2001}({van Dokkum} \& {Stanford} 2001).  Most likely, discs in
early-type galaxies are the result of remaining gas settling after major merging
events between two late-type galaxies \nocite{naab2001}({Naab} \& {Burkert}
2001).  Another possible formation scenario for discs involves slow accretion
from many minor merging events.  Debris from minor mergers, or high velocity
clouds in the halo of the galaxy, could rain down onto the disc
\nocite{benjamin1997,putman2002}({Benjamin} \& {Danly} 1997; {Putman} {et~al.}
2002).  However, such a scenario would lead to thicker, less pronounced disc
structures.  Since late-type galaxies exhibit clear disc structures, this
process is unlikely to be very important in the formation of typical spiral
structures.

In all these scenarios, the obvious question arises as to whether disc
structures are exclusively found in late-type galaxies.  Observationally, that
is clearly not the case.  S0 galaxies in the local universe possess disc
components, and even local ellipticals show evidence of disc structures and dust
lanes when studied in detail \nocite{rest2001,tran2001}({Rest} {et~al.}  2001;
{Tran} {et~al.}  2001).  The isophotes of ellipticals are usually classified
into ``boxy'' and ``discy'', types.  Pointed, discy isophotes may well be caused
by the presence of discs.  Even in the case when the isophotes are nearly round
and show no sign of ``discyness'', kinematic studies suggest that it is still
possible to hide a disc in such galaxies \nocite{romanowsky1997}({Romanowsky} \&
{Kochanek} 1997).  

There is also some observational evidence for discs in ellipticals at
cosmological redshifts; \nocite{kelson2000}{Kelson} {et~al.}  (2000) found that
many early-type galaxies in CL 1358+62 at $z=0.33$ possess evidence for the
presence of exponential discs, suggesting discs are present in a large fraction
of early-type galaxies, at least in clusters.  Since there is no clear
understanding yet on how these discs form, and no systematic observational
investigation into the abundance and properties of discs in early-type field
galaxies has been carried out, it is difficult to make precise statements about
the abundance of discs in field ellipticals.  However, some recent observations
using the \emph{Hubble Space Telescope} (\emph{HST}) show that a fraction of at
least $\sim50$\,per cent of field ellipticals contain discs of sizes spanning a
range of at least an order of magnitude, from small nuclear discs to large discs
extending beyond the bulge \nocite{rest2001}({Rest} {et~al.}  2001).  In
addition, kinematic studies using the \emph{SAURON} integral field spectrograph,
also suggest that a high fraction of $\sim50$\,per cent of field ellipticals
contain discs of various sizes \nocite{deZeeuw2002}({de Zeeuw} {et~al.}  2002).

Most lens galaxies known to date are early-type galaxies.  The main reason for
this is that early-type galaxies tend to be the most massive and hence have a
larger cross section for gravitational lensing.  However, as shown by Blain,
M\"oller \& Maller (1999) spiral galaxy lenses are also predicted to contribute
significantly to the total lensing cross section.  This is mainly due to the
high surface mass density of the thin disc component when viewed edge on
(Maller, Flores \& Primack, 1997; M\"oller \& Blain, 1998).  As shown in those
papers, the disc components of late-type lenses can make important contributions
to the lensing potential and change the generic properties of the lens systems.
The discovery of spiral lens systems in the CLASS lensing survey, e.g.
B1600+434 \nocite{koopmans1998}({Koopmans}, {de Bruyn} \& {Jackson} 1998),
confirm that this is the case.  However, early-type lensing galaxies have so far
nearly always been modelled with a simple one-component model, possibly
including external shear (Keeton, Kochanek \& Seljak, 1997; Leh\'ar et al.,
2000).

In this paper we determine the lensing properties of early-type galaxies that
contain realistic discs with stellar masses of up to $\sim10$\,per cent of that
of the bulge.  We will determine the lensing properties of early-type lenses and
show how discs in ellipticals could explain some properties of observed lens
systems that have not yet been modelled successfully.  We concentrate on three
main questions:  
\begin{enumerate} 
\item statistical lensing properties:  how do
disc structures in early-type lensing galaxies affect the statistical properties
of a sample of early-type lenses?  We calculate magnification cross sections and
the expected ratio of quad:  two-image systems.  In particular, we investigate
whether discs in early-type lenses may explain the high ratio of quad image
systems, which is still an unsolved enigma in gravitational lensing
(\nocite{rusin2001}{Rusin} \& {Tegmark}, 2001, but see \nocite{chae2002}{Chae}, 2002).  

\item magnification ratios:  can disc structures explain some or all of the
unusual magnification ratios observed in some early-type lens systems?  To date,
these deviant magnification ratios have been interpreted as providing strong
evidence in favour of the presence of compact halo substructure as predicted by
current cold dark matter (CDM) simulations \nocite{metcalf2001}(e.g.  {Metcalf}
\& {Madau} 2001).  \item time delays:  how would disc structures affect the
measured time delays in early-type lens systems?  Currently there seems to be a
strong disagreement between the value of $H_0$ as found in the \emph{HST} Key
Project \nocite{freedman2001}({Freedman} {et~al.}  2001) and measurements of
$H_0$ from gravitational lensing \nocite{kochanek2003}({Kochanek} 2003).  Some
local estimates of $H_0$ (e.g.\nocite{saha2001}{Saha} {et~al} 2001) also favour
lower values.  We investigate whether disc structures in gravitational lens
systems may reduce this discrepancy.  \end{enumerate} We start in
\S\ref{surbripro} by discussing the appearance of the model lens galaxies and
show that the manifestations of the presence of a disc are similar to those
observed in some ellipticals.  In \S\ref{models} we present the parametric lens
models used in this paper.  In \S\ref{properties} we describe some general
and statistical lensing properties of early-type galaxies that include disc
structures.  In \S\ref{magratios} we show how discs in ellipticals could explain
some of the observed magnification ratios and in \S\ref{hubble} we calculate how
the presence of discs in ellipticals would affect estimates of $H_0$ from
lensing.  We conclude with a summary in \S\ref{conclusions}.

Throughout this paper we assume a Friedmann-Robertson-Walker Universe with
$\Omega_{\mathrm{M}}=0.3$, $\Omega_{\mathrm{\Lambda}}=0.7$ and
$H_0=50\,\mathrm{km}\,\mathrm{s}^{-1}\,\mathrm{Mpc}^{-1}$.  The choice for this
value of $H_0$ is motivated due to limitations of the original lensing code.
Subsequently, the code has been extended to allow for any value of $H_0$, as
necessary for the calculations in \S\ref{hubble}, but we did not recalculate the
results of \S\ref{models}-\ref{magratios}, as these calculations are very
time-consuming and a change in $H_0$ has little effect on the results.  A value
of $H_0=70\,\mathrm{km}\,\mathrm{s}^{-1}\,\mathrm{Mpc}^{-1}$, is probably more
in line with current observational constraints, as given for example by the
\emph{HST} Key Project \nocite{freedman2001}({Freedman} {et~al.}  2001) and the WMAP results \nocite{bennett2003}({Bennett} {et~al.} 2003).


\section{Surface Brightness Profiles}
\label{surbripro}

The existence of disc components with large scale-lengths in luminous elliptical
galaxies has been known for some time \nocite{bend1989, rh1990,
rix1991}({Bender} {et~al.}  1989; {Rix} \& {White} 1990; {Rix} 1991).  Recent
theoretical work \nocite{naab2001, barn2002}({Naab} \& {Burkert} 2001; {Barnes}
2002) has begun to address the origin of such components in bulge-dominated
galaxies.  However, the ability to reliably determine the prevalence of discs in
strongly bulge-dominated systems has only been possible following the
development of full two-dimensional decomposition techniques
\nocite{byun1995}({Byun} \& {Freeman} 1995).  Given data of adequate
signal-to-noise ratio and angular resolution, two-dimensional decomposition
methods are capable of identifying discs in systems with bulge fractions (B/T)
as high as B/T=0.95.  A significant quantity of {\it HST} imaging, obtained
using the Wide Field Planetary Camera 2 (WFPC2), of luminous bulge-dominated
galaxies in intermediate redshift clusters is well-suited to such decompositions
but little work appears to have been undertaken.  Exceptions are the
investigations of the $z=0.33$ cluster CL1358+62 by \nocite{kelson2000}{Kelson}
{et~al.}  (2000) and \nocite{tran2003}{Tran} {et~al.}  (2003).  Fig.\,4 of
Kelson et al.  provides a particularly striking demonstration of the presence of
discs, even among galaxies with conventional `E' morphological classifications.

In order to parameterize the  light profile of ellipticals, a commonly
used set of parameters  are the half-light radius $r_{\mathrm{h}}$ and
the surface brightness at  the half light radius $I_{\mathrm{h}}$. The
surface brightness  of a de-Vaucouleurs bulge  can then be  written as
\nocite{binney1987}({Binney} \& {Tremaine} 1987),
\begin{equation}
I(r)=I_{\mathrm{h}}\exp{\left\{-7.67[(r/r_{\mathrm{h}})^{(1/4)}-1]\right\}}.
\label{eq.light1}
\end{equation}
The parameterization of the surface brightness distribution for
an exponential disc, in terms of the central
surface brightness, $I_{\mathrm{d}}$, and scale--length, $r_{\mathrm{d}}$, 
can be written

\begin{equation}
I(r)=I_{\mathrm{d}}\exp{-(r/r_{\mathrm{d}})}.
\label{eq.light2}
\end{equation}

The bulge fraction, B/T, for a galaxy parameterized by such bulge and disc
components is then

\begin{equation}
{\rm {B \over T}} = { I_{\mathrm{h}} r_{\mathrm{h}}^2 \over
I_{\mathrm{h}} r_{\mathrm{h}}^2 + 0.28 I_{\mathrm{d}}
r_{\mathrm{d}}^2}
\label{eq.iso}
\end{equation}
where $I_{\mathrm{h}}$ and $r_{\mathrm{h}}$, $I_{\mathrm{d}}$ and
$r_{\mathrm{d}}$ are as defined above.  

Table 2 in Kelson et al. gives surface brightnesses and half-light-radii for
bulge-only and bulge-plus-disc fits to galaxies in CL1358+62.  Note that Kelson
et al.  specify the properties of the disc component in terms of half-light
radius of the disc, $r_{\mathrm{hd}}=1.688r_{\mathrm{d}}$, and the mean surface brightness within
the half-light radius, $<I_{\mathrm{hd}}>$.  Using this parameterization the B/T
ratio may be calculated by replacing the term $0.28 I_{\mathrm{d}}
r_{\mathrm{d}}^2$ by $<I_{\mathrm{hd}}> r_{\mathrm{hd}}^2$.  

For the investigation of the general and statistical lensing properties of discy
ellipticals, we use the full sample of galaxy models in Kelson's catalogue.
However, our main results, as presented in \S\,\ref{magratios} and
\S\,\ref{hubble} involve time-consuming calculations and are therefore based on
the parameters describing a single model galaxy.  The presence of many galaxies
with bulge and disc components with comparable scalelengths in CL1358+62 is
clear from the form of the residuals shown in Figure 4 of Kelson et al.  It is
worth stressing that the presence of even highly inclined (almost edge-on)
discs, with scale-lengths similar to those identified by Kelson et al., making
up less than $\sim 10$\,per cent of the total light are hard to discern directly
in images.  Plots showing the residuals obtained by fitting a bulge-only model
are, however, far more revealing.  The presence of face-on discs is revealed by
characteristic donut-like residuals, while edge-on discs appear as often quite
prominent inclined structures.  Notwithstanding the evidence provided by Kelson
et al.  in support of the accuracy of their bulge-disc compositions we have
chosen to be conservative by reducing the disc contribution to each galaxy in
the Kelson et al.  catalogue by half.  Thus, our investigation of the
statistical lensing properties is based on the properties of the galaxies listed
in Table 2 of Kelson et al.  but with a reduction in the disc contribution of a
factor 2.  For the individual model galaxy used in our calculations we have
chosen galaxy no.\,242 (Table\,\ref{table.short}).  Adopting our rescaling of
the disc contribution, galaxy no.\,242 has B/T=0.95, c.f., B/T=0.91 in Kelson et
al. Galaxy no.\,242 has a total mass, B/T ratio and bulge scale-length close to
the average value for the `E' galaxies in Kelson et al's catalogue.  In our
simulations the ellipticity of the bulge is set to $e=0.4$ and the inclination
of the disc is $i=75\,\deg$.

\begin{figure*}
\epsfig{file=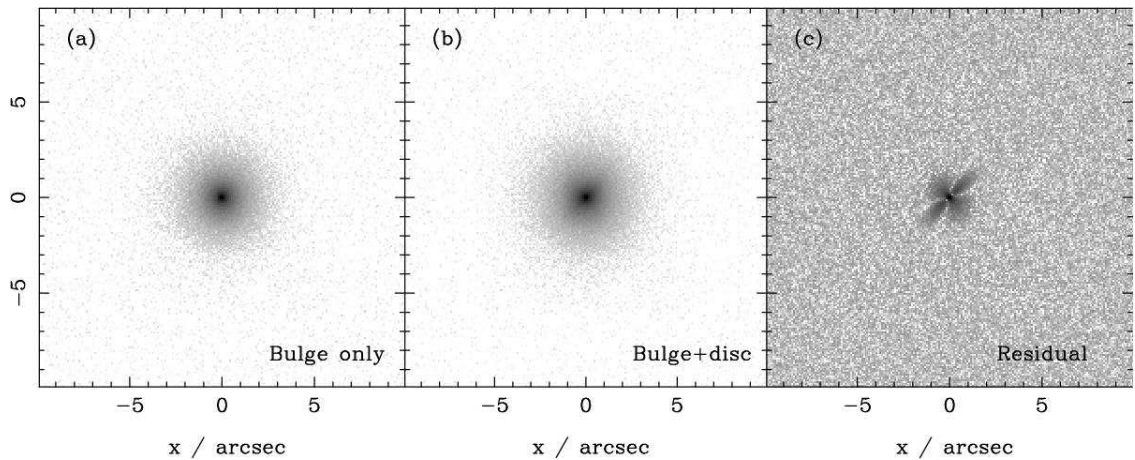,width=6.0cm,angle=-90}\\

\caption{Simulated images of a typical model galaxy at redshift $z=0.33$.  The
galaxy model parameters correspond to our modified representation of galaxy
no.\,242 in Kelson's catalogue (Table 1).  Panel (a) shows the image derived
using a bulge-only model.  Panel (b) shows the bulge-plus-disc model,
normalized to have the same total counts as the bulge--only model.  The disc
component of the model is inclined at $70\,\deg$ to the line of sight and oriented
at an angle of $45\,\deg$, to the $x$ axis.  Panel (c) shows the residual
following subtraction of the bulge-only image from the bulge-plus-disc image.  The
pixel scale, $0.1\arcseca$ per pixel, and point spread function have been chosen
to approximate images obtained using the WFPC2 on {\it HST}.  The signal-to-noise
ratio of the synthetic images is significantly higher than that typical of {\it
HST} imaging data of the type analysed by Kelson et al.  (2000).}

\label{isophotes.fig}
\end{figure*}

Fig.\,\ref{isophotes.fig} shows images based on our adopted parameterization of
galaxy no.\,242 with B/T=0.95.  Fig.\,\ref{isophotes.fig}a is an image based on
the bulge-only model surface-brightness profile fit (parameters from Cols:  3 \& 4
in Table 1), Fig.\,\ref{isophotes.fig}b shows a realization of the bulge-plus-disc
surface-brightness profile (parameters from Cols:  7 \& 8 (bulge) and Cols:  10 \&
11 (disc) in Table 1).  The disc is at high inclination, $i=70\,\deg$ to the line
of sight ($i=90\,\deg$ corresponding to edge-on) and oriented at an angle of
$45\,\deg$ with respect to the x-axis.  The presence of the disc is barely
discernible.  Fig.\,\ref{isophotes.fig}c shows the residual image derived from
subtracting Fig.\,\ref{isophotes.fig}a from Fig.\,\ref{isophotes.fig}b (both
normalized to possess the same total counts).  The images may be compared to those
in Figs 1 and 4 of Kelson et al..  In Kelson et al., galaxy no.\,242 shows the
donut-like residual characteristic of the presence of a close to face-on disc.  By
contrast, in Fig.\,\ref{isophotes.fig}c, the disc appears prominently because of
the high inclination angle of $i=70\,\deg$ to the line-of-sight.

Following submission of the original version of this paper an analysis of the
photometric properties of galaxies in CL~1358+62 undertaken by Tran et al.
(2003) appeared.  Comparison of both the B/T values and the scale-lengths of the
light distributions for the bulge-dominated galaxies in common between Kelson et
al.  and Tran et al.  show a good correlation over the full range of B/T.
However, for large values of B/T the contribution of the disc components are
much larger in the Tran et al.  analysis.  Less satisfactory still is the
comparison between the scale-lengths of the disc components found among the
bulge-dominated galaxies.  The disc scale-lengths found by Tran et al.  are
significantly larger, explaining, at least in part, the systematically smaller
B/T ratios found for the bulge-dominated systems.  Tran et al.  present the
results of a large number of Monte-Carlo simulations designed to quantify the
accuracy of the decompositions into bulge and disc components.  No reference is
made to the strong systematic disagreement between the statistics of the disc
scale-lengths found in the two studies.  Tran et al.  are confident that their
simulations demonstrate the ability of the software employed to obtain accurate
bulge-disc decompositions. However, the parameter ranges employed in the 
simulations appear to be determined from the results of the actual decompositions 
applied to the real data. 
Tran et al. find no discs with scale-lengths comparable to 
or smaller than those of the bulge components of the galaxies. Thus,
it is not clear that the ability of Tran et al.'s analysis to correctly
parameterize combinations of bulge and disc profiles with parameters found 
by Kelson et al. has been adequately tested. In the circumstances we have
decided to retain the very conservative distribution of B/T ratios,
with associated values of component scale-lengths, based
on the distribution advocated by Kelson et al.

\section{The lens models}
\label{models}

In order to study the lensing properties of early-type galaxies including discs,
a working model of the mass distribution is needed.  Unfortunately, little is
known about the detailed properties of early-type galaxy discs and how they
depend on morphology and environment.  From past studies it is clear that discs
in ellipticals are abundant and that the disc sizes and magnitudes cover the
whole range from compact and luminous to extended and faint
\nocite{rest2001,scorza1995,scorza1998}({Rest} {et~al.}  2001; {Scorza} \&
{Bender} 1995; {Scorza} {et~al.}  1998).  Most of the available information is
based on observations of individual galaxies; there has not yet been a
systematic high-resolution study of the morphologies of a complete, unbiased
sample of early-type galaxies.  In order to make meaningful predictions about
the lensing properties of early-type lensing galaxies including disc components,
we base our models on the light profiles obtained in \nocite{kelson2000}{Kelson}
{et~al.}  (2000).  All galaxies in Kelson's sample are at a redshift of
$z=0.33$, which corresponds roughly to the redshift at which lensing is most
efficient; few lenses are expected with redshifts less than $z<0.1$.

\subsection{Parameterization of early-type lenses with discs}
We   model  the  elliptical  lensing  galaxies  with two-components:
an elliptical  bulge with  a de-Vaucouleurs $r^{1/4}$ profile and an
exponential thin  disc.  

Under the usual assumption of a thin lens,  the lensing properties
depend solely on the  surface mass density profile. For the bulge
component this is given by
\begin{equation}
\Sigma_{\mathrm{b}}(r)=\Sigma_{\mathrm{b}}\exp{(-r'/r_{\mathrm{b}})^{1/4}},
\label{bulgesig}
\end{equation}
where $\Sigma_{\mathrm b}$ is the central  surface mass density of the bulge, $r_{\mathrm{b}}$ is the bulge scale-length, and
\begin{equation}
r'=\sqrt{x^2+y^2(1-e)^2}.
\end{equation}
Here $r=\sqrt{x^2+y^2}$, and $e$ is the ellipticity of the bulge. The disc, inclined at an angle $i$ to the line of sight, can be parameterized in a
similar way, to give
\begin{equation}
\Sigma_{\mathrm{d}}(r)=\Sigma_{\mathrm d}\exp{(-r'/r_{\mathrm{d}})},
\label{discsig}
\end{equation}
where $\Sigma_{\mathrm d}$ is the central surface mass  density of the disc, 
$r_{\mathrm{d}}$ is the scale-length of the disc and
\begin{equation}
r'=\sqrt{x^2+y^2[1-\cos{(i)}]^2}.
\end{equation}

We convert the light profiles as given in equations\,(\ref{eq.light1}) \&
\,(\ref{eq.light2}) to models of the mass distribution assuming a constant mass-to-light
ratio, $\Gamma$.  The relation between the parameters in
equations\,(\ref{eq.light1}) \&\,(\ref{eq.light2}) and those in
equations\,(\ref{bulgesig})\,\&\,(\ref{discsig}), is then
$r_{\mathrm{h}}=7.676^4r_{\mathrm{b}}$ and
$I_{\mathrm{h}}=\Gamma^{-1}\Sigma_{\mathrm{b}}\exp{(-7.676)}$ for the bulge
component and
$I_{\mathrm{d}}=\Gamma^{-1}\Sigma_{\mathrm{d}}$ for the disc
component.  We will refer to $r_{\mathrm{h}}$ and $r_{\mathrm{hd}}=1.688r_{\mathrm{d}}$ as the
`effective radius' or `half-light radius' of bulge and disc, respectively.
Assuming randomly oriented discs, the average inclination is about $70\,\deg$.
Unless otherwise stated, the discs in our bulge-plus-disc models have this
inclination.

\subsection{Possible caveats of the model}

The photometric profile decompositions of galaxies in CL 1358+62 provide an
observational motivation for modelling early-type galaxies including disc
structures.  However, galaxies in cluster environments might exhibit more
pronounced discs than corresponding galaxies in the field.  If this is indeed
the case, and simulations of \nocite{naab2001}{Naab} \& {Burkert} (2001) suggest
otherwise, the discs in our model lensing galaxies could be too pronounced
compared to those expected in observed early-type lens systems, many of which
are in lower-density field environments.  A comparison of the disc sizes
obtained from a sample of mostly field ellipticals by
\nocite{scorza1995}{Scorza} \& {Bender} (1995) with the sizes of discs in E's
and S0's found in \nocite{kelson2000}{Kelson} {et~al.}  (2000) shows that the
discs from both samples are similar and range in disc half-light radii from
$0.1\,\mathrm{kpc}$ to $1\,\mathrm{kpc}$.  This could indicate that the
environment may influence the relative number of galaxies as a function of
morphology, but has little effect on the properties of discs in galaxies of a
given morphological type.
In any case, we are not concerned here with accurate modelling of the 
statistical incidence of discs, but rather aim to show how discs with sizes 
and masses consistent with current observations influence the 
gravitational lensing properties of early-type lens systems. 

There is a second possible caveat with the mass model we use here: we assume that mass follows light and do not include a dark matter halo.
As we will discuss in more detail below, the main effect of inclined, thin disc
components on the lensing properties of galaxies arises from the strong asymmetry
that is introduced in the central regions of the galaxies due to the disc.  We
therefore expect that the precise form of the spherical or nearly-spherical
components in which the disc is embedded, that is, the bulge and halo, will not
change the effect of the added disc component drastically.  We have checked this
explicitly for the mass model used to derive the results in \S\ref{magratios} and
\S\ref{hubble}, by adding a dark matter halo that contains 2/3 of the total mass
and scaling the mass of the bulge and disc so that the Einstein radius is the same
as for the model without a dark-matter halo.  Estimates of the mass-to-light ratio
of early type galaxies range from $M/L\sim2$ in the K-band \nocite{moriondo1998}(e.g.
Moriondo, Giovanardi \& Hunt 1998) to about $M/L\sim10$ in the V-band
\nocite{natarajan1998}(Natarajan et al.  1998).  Many lens systems are consistent
with lower mass-to-light ratios within an Einstein radius of about $M/L\sim2$ in V
\nocite{koopmans2003}(Koopmans \& Treu 2003), motivating our choice of a dark halo
containing 2/3 of the total mass.  When including such a spherical dark halo we
find that the inner caustic shrinks in size, and so quad images become less
likely, but that the magnification ratios and time delays of images of sources
that are quadruply imaged do not change substantially. 

\subsection{Surface density and mass profiles of lens models} 
\begin{figure}
\epsfig{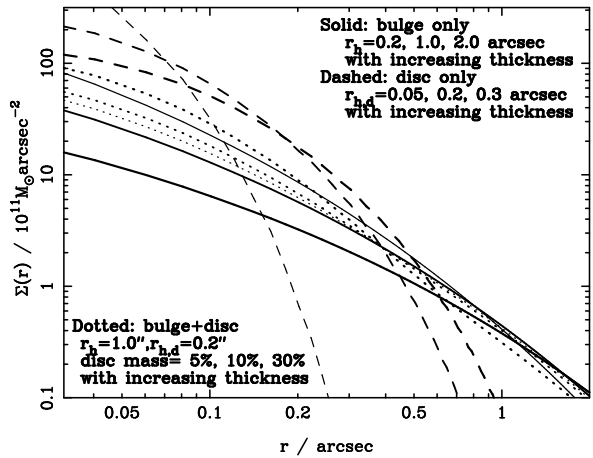}\\
\caption{Radial surface mass density profiles for representative
models of the galaxies in the cluster CL 1358+62. The different lines
show the surface mass density as a function of radius for different
model parameters that are representative for the majority of all
galaxies in the sample. All curves are normalized so that the total
mass is $10^{11}M_{\odot}$.  The ellipticity of the bulge is $e=0.4$
and the inclination angle of the disc is $i=70\,\deg$.  The surface
mass density differs by less than $10$\,per cent between models that include a
disc with $M_{\mathrm{d}}=0.1M_{\mathrm{b}}$ and models without a
disc. At small radii, discs increase the surface mass density, due to
their smaller scale lengths. At the cluster redshift of $z=0.33$,
$1\arcseca$ corresponds to $6.6\,\mathrm{kpc}$ for the cosmology we
adopt in this paper.}
\label{fig.radial}
\end{figure}

We show  the radial surface mass density profiles for  some typical
cases in Fig.\,\ref{fig.radial}. The solid lines show the density
profiles of three bulge-only models with effective radii of
0.2, 1.0 and 2.0\,arcsec. At the cluster redshift
of $z=0.33$ these values correspond to $1.3$, $6.6$ and
$13.3\,\mathrm{kpc}$ respectively. These values cover the range of
half-light radii of the bulge-only fits to the light profiles of the
galaxies in CL 1358+62, which lie between 0.2 and
$4.9\arcseca$ with only two galaxies having half-light radii larger than
$1.6\arcseca$. The dashed lines in Fig.\,\ref{fig.radial} show the
surface mass densities for discs of effective radii of 0.05,
0.2 and $0.3\arcseca$. The discs are assumed to be inclined at
$70\,\deg$ to the line of sight.  Most of the bulge-plus-disc fits to the
light profile of the elliptical galaxies in the sample have discs 
with effective radii less than $0.3\arcseca$, corresponding to $2\,\mathrm{kpc}$
at $z=0.33$. The disc components of the later type galaxies are
usually larger, with effective radii of up to $1\arcseca$. Comparison of
the relative bulge and disc contributions shows that the disc
components contribute significantly to the surface mass density within
about $0.5\arcseca$, or $3.3\,\mathrm{kpc}$. The dotted
lines show the total surface mass density for a typical bulge-plus-disc
fit, but with different B/T mass ratios. It is clear that the radial
profiles for models that include a  disc of less  than $\sim5$\,per cent of
the total mass are very similar to bulge-only models. For B/T mass
ratios of $0.9$ or less, that is for most of the S0/Sa or later type
galaxies in Kelson et al., the surface mass density in the inner
$0.2\arcseca$ differs by more than $10$\,per cent.

\begin{figure}
\epsfig{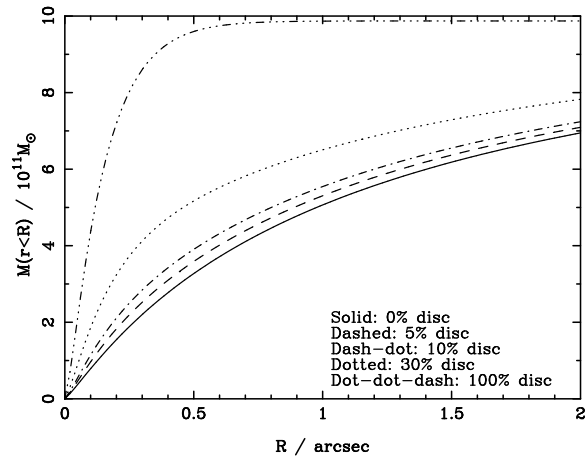}\\
\caption{Radial mass profiles for a number of galaxy models. The
different lines show the enclosed mass as a function of radius for
different model parameters. The curves are normalized so that
$\lim_{R\to\infty} M(R)=10^{11}M_{\odot}$. The effective radii for bulge
and disc are $r_{\mathrm{hd}}=r_{\mathrm{h}}=0.2\,\arcseca$, which
corresponds to $1.3\,\mathrm{kpc}$ at the cluster redshift of
$z=0.33$.  The ellipticity of the bulge is $e=0.4$ and the inclination
angle of the disc is $75\,\deg$.  The enclosed mass at distances of
$\sim1\arcseca$ differs by less than $10\%$ between models that include
a disc containing $10\%$ of the mass and models without a disc. Note
that, as in the previous figure, the most significant contribution of
the disc is in the inner regions.}
\label{fig.mass}
\end{figure}
Gravitational   lensing   measures   directly   the   projected   mass
$M_{\mathrm{enc}}$  contained  inside  a  circular  region  of  radius
$R$.  In Fig.\,\ref{fig.mass} we  show $M_{\mathrm{enc}}$ as a
function of  radius $R$. All the models  shown are normalized
to the same total  mass, but  the B/T mass ratio  varies. The 
effective radius of disc and bulge, ellipticity  of bulge  and inclination
angle of disc are $r_{\mathrm{hd}}=0.2\,\arcseca$,
$r_{\mathrm{h}}=0.2\,\arcseca$, $e=0.4$ and $i=75\,\deg$ respectively
and are the same for all models.  These values correspond to the
typical values for an elliptical galaxy in CL 1358+62. At distances of
$R\sim1\arcseca$ the  enclosed mass for typical bulge-plus-disc models with
B/T$>0.9$ is  only a few  percent larger than  the enclosed mass for
bulge-only models. None of the discs in the light profile fits to the
early-type galaxies in CL 1358+62 contain more than $\sim10$\,per cent of the
light (cf.  Table A1), and so, if the mass-to-light ratio is constant, then 
disc structures are expected to contain at most $10$\,per cent of the
mass. The effect of the disc on the maximum image separations or the
radius of the Einstein ring will be small in those cases. More massive
discs with a mass of $M_{\mathrm{d}}\sim0.3M_{\mathrm{b}}$, contribute
significantly to the mass in the inner region of the galaxy and
influence the mass contained within the expected Einstein radius more
strongly.

The properties of all the elliptical galaxies from the Kelson et al. (2000)
catalogue with the reduced disc contribution adopted for our simulations are
summarized in Table\,\ref{table.short}.  The table lists both the values for the
best fitting bulge-only models and the values for the best bulge-plus-disc fits.
The two parameters $f_{\mathrm{e}}$ and $f_{\mathrm{t}}$ give the B/T mass ratio
as measured at the Einstein radius $R_{\mathrm{E}}$ and at $R\to\infty$
respectively.  The Einstein radius is calculated taking account of the
ellipticity of the bulge and the inclination of the disc.  Note that the
Einstein radius does not depend strongly on whether a disc is included or not:
there is no systematic dependence of the Einstein radius on the presence of a
disc.

\begin{table*}
\caption{Elliptical galaxies in CL 1358+62. The values for
$<I_{\mathrm{h}}>$, $r_{\mathrm{h}}$ and
$<I_{\mathrm{hd}}>$ are taken directly from Kelson et al. (2000). The disc scale length $r_{\mathrm{d}}$ is related to the half-light radius as given in Kelson et al. by $r_{\mathrm{d}}=1.688r_{\mathrm{hd}}$. In a
cosmology with $\Omega_{\Lambda}=0.7$,
$\Omega_{\mathrm{M}}=0.3$ and
$H_0=50\,\mathrm{km}\,\mathrm{s}^{-1}\,\mathrm{Mpc}^{-1}$, $1\arcseca$
corresponds to $6.6\,\mathrm{kpc}$ at the cluster redshift of
$z=0.33$.  A larger table that includes all the galaxies in CL 1358+62
investigated by Kelson et al. (2000), is given in the Appendix.}
\begin{tabular}{l|r|r|r|r|r||r|r|r|r|r|r|r|r|r}\hline
 & & \multicolumn{1}{l|}{$\longleftarrow$} & \multicolumn{2}{c|}{Bulge only} & $\longrightarrow$ & \multicolumn{1}{l|}{$\longleftarrow$} & \multicolumn{7}{c|}{Bulge and disc} & $\longrightarrow$\\ \hline 
No. & Type & $<I_{\mathrm{h}}>$ & $r_{\mathrm{h}}$ & $\Sigma_{\mathrm{b}}$ & $R_{\mathrm{e}}$ & $<I_{\mathrm{h}}>$ & $r_{\mathrm{h}}$ & $\Sigma_{\mathrm{h}}$ & $<I_{\mathrm{hd}}>$ & $r_{\mathrm{d}}$ & $\Sigma_{\mathrm{d}}$ & $R_{\mathrm{e}}$ & $f_{\mathrm{e}}$ & $f_{\mathrm{t}}$\\ \hline
212 & E & 20.780 & 0.683 & 0.488 & 1.220 & 21.170 & 0.781 & 0.340 & 18.030 & 0.031 & 6.138 & 1.160 & 0.808 & 0.960\\
242 & E & 21.920 & 1.529 & 0.171 & 1.360 & 22.420 & 1.825 & 0.108 & 20.720 & 0.156 & 0.515 & 1.270 & 0.806 & 0.950\\
256 & E & 20.880 & 1.380 & 0.445 & 2.330 & 21.200 & 1.551 & 0.331 & 18.000 & 0.053 & 6.310 & 2.260 & 0.836 & 0.967\\
303 & E & 20.590 & 0.638 & 0.581 & 1.270 & 21.110 & 0.774 & 0.360 & 18.890 & 0.054 & 2.780 & 1.210 & 0.787 & 0.947\\
360 & E & 20.220 & 0.341 & 0.817 & 2.176 & 21.660 & 0.594 & 0.217 & 17.620 & 0.030 & 8.954 & 1.000 & 0.565 & 0.867\\
375 & E & 22.860 & 4.979 & 0.072 & 2.650 & 23.010 & 5.267 & 0.063 & 19.220 & 0.096 & 2.051 & 2.870 & 0.800 & 0.984\\
409 & E & 21.270 & 0.498 & 0.310 & 2.176 & 21.490 & 0.542 & 0.254 & 22.600 & 0.132 & 0.091 & 0.660 & 0.959 & 0.969\\
412 & E & 21.390 & 0.767 & 0.278 & 2.176 & 22.320 & 1.091 & 0.118 & 17.660 & 0.033 & 8.630 & 0.870 & 0.502 & 0.908\\
531 & E & 21.260 & 1.549 & 0.313 & 2.090 & 21.710 & 1.830 & 0.207 & 19.300 & 0.108 & 1.905 & 1.980 & 0.772 & 0.954\\
534 & E & 21.210 & 0.620 & 0.328 & 0.860 & 22.170 & 0.878 & 0.136 & 16.680 & 0.019 & 1.281 & 0.770 & 0.489 & 0.901\\
536 & E & 21.180 & 1.266 & 0.337 & 1.790 & 21.520 & 1.433 & 0.247 & 18.570 & 0.057 & 3.733 & 1.720 & 0.807 & 0.965\\
\hline
\end{tabular}

\label{table.short}
\end{table*}

\section{Lensing Properties}
\label{properties}
The lensing properties  of discs in spiral galaxies  have been studied
by \nocite{maller1997,moller1998,blain1999a}{Maller}, {Flores} \& 
{Primack} (1997), {M{\"o}ller} \& {Blain} (1998) and 
{Blain}, {M{\"o}ller}, \&  {Maller} (1999).  In  those
studies  it  was  found that  thin  exponential  discs  can influence
the statistical lensing properties even if they only contain $10$\,per cent of
the total mass provided that they are inclined by more than
about $70\,\deg$ to the line of sight. In particular, the  cross
section  for  image  geometries with  4  or more  magnified images,
and the cross section for high magnifications of 10 or more is
increased considerably.   These  results   are   generic  and   hold
qualitatively for discs in early-type galaxies. However, the scale
lengths and masses of the disc components in early-type galaxies are likely to
be smaller, relative to the bulge component, and so the
effect of  the  disc on  their statistical  lensing  properties  is
expected to  be  less. In this section, we describe the general and 
statistical lensing properties of discy early-type galaxies. 

\subsection{General lensing properties}
The gravitational potential of a mass distribution between an observer
at $z=0$ and a source at $z=z_{\mathrm{s}}$ deflects the light of the
source, at position $\vec{\beta}$ on the source plane, so that an
image is observed at position $\vec{\theta}$ on the lens plane, where
\begin{equation}
\vec{\theta}=\vec{\beta}+\frac{D_{\mathrm{LS}}}{D_{\mathrm{OS}}}\vec{\alpha}.
\end{equation}
The deflection angle of a light ray $\alpha$ that passes the lens at
an angular position $\vec{\theta}$ relative to the lens centre, is
given by
\begin{equation}
\vec{\alpha}(\vec{\theta})=\frac{D_{\mathrm{OS}}}{\pi
D_{\mathrm{LS}}}\int\kappa(\vec{\theta'})\frac{\vec{\theta}-\vec{\theta'}}{\left|\vec{\theta}-\vec{\theta'}\right|^2}\,d^2\theta'.
\label{defleq}
\end{equation}
where $D_{\mathrm{OS}}$ and $D_{\mathrm{LS}}$ are the angular diameter
distances from observer to source and lens to source respectively.
The dimensionless quantity $\kappa$ is defined as
$\kappa=\Sigma/\Sigma_{\mathrm{c}}$ where
\begin{equation}
\Sigma_{\mathrm{c}}=\frac{c^2D_{\mathrm{OS}}}{4\pi
GD_{\mathrm{OL}}D_{\mathrm{LS}}}
\end{equation}
is the critical surface mass density. These equations show explicitly
why thin disc components may be important, even if they contain only a
small fraction of the total mass: it is the projected surface mass
density that enters equation\,(\ref{defleq}) and this can be very high
for thin edge-on discs.

The numerical ray-tracing routines that  we used to obtain the results
in  this paper are  described  in  detail  in
\nocite{moller1998}{M{\"o}ller} \& {Blain} (1998, 2001). The routines
were  used to  generate magnification  maps  on the  source plane  and
determine  the cross  sections  for high  magnifications and  multiple
imaging.
\begin{figure}
\epsfig{file=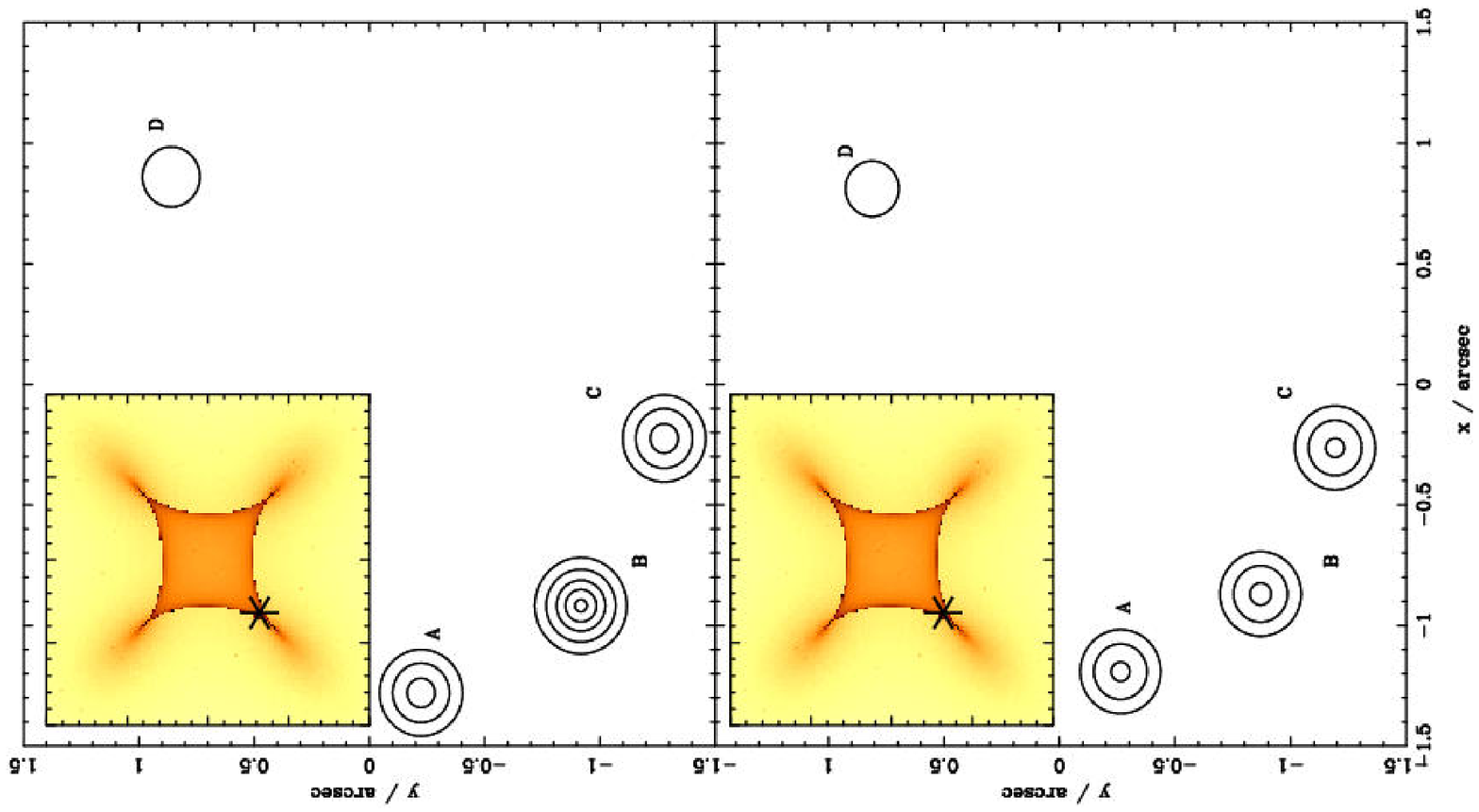,width=14.0cm,angle=-90}\\
\caption{Lensing of a point source by an elliptical galaxy with and
without a disc. The contours show the images of a point source at
$z_{\mathrm{s}}=1$ that lies behind a lens galaxy with the same
properties as galaxy no.\,242 in Table\,\ref{table.short}, convolved
with a Gaussian beam of $0.1\arcseca$ FWHM. The contour levels are 0.5,
3, 10, 20, and 30 in arbitrary flux units. The top panel shows the
resulting images for a bulge-only model. The bottom panel shows
the images for a bulge-plus-disc model, where the disc is inclined at
$i=75\,\mathrm{deg}$ to the line of sight. The major axis of both
bulge and disc is at $45\deg$ to the x-axis. The insets show the
respective $0.2\arcseca\times0.2\arcseca$ magnification maps on the 
source plane. 
The grey-scale is logarithmic
and ranges from $\mu\sim2$ (white) to $\mu\sim200$ (black). The source
position is marked by a star in the inset, and is
slightly different in the two panels, to keep the
image geometry very similar and the total source magnification 
roughly the same in both cases.}
\label{properties.sing.fig}
\end{figure}

To illustrate the possible effect of disc components in a lens system,
Fig.\,\ref{properties.sing.fig} shows the images of a background point
source at a redshift  of $z_{\mathrm{s}}=1.0$ that is lensed  by a
foreground elliptical galaxy at redshift  $z_{\mathrm{l}}=0.3$. The
images are shown for the bulge-plus-disc and pure-bulge model discussed in \S\,\ref{surbripro}.
The source position is adjusted slightly, by less than
$0.01\arcseca$, so that both the total magnification of the source and the
image geometries are very similar. Note the change in the
relative fluxes of the images.  
The insets
show the magnification maps on the source plane, $\mu(\vec{\theta})$
for both cases; the  disc increases the area enclosed by the caustic
only slightly. 
Comparing the magnification maps on
the source plane as shown in the insets of
Fig.\,\ref{properties.sing.fig} with those of disc lenses
shown in Fig.\,3 in \nocite{moller1998}{M{\"o}ller} \& {Blain} (1998), makes clear that the
effect of a disc component in early-type lensing galaxies is much less
pronounced.  However, there are differences in the magnification ratios which we will investigate more in \S\ref{magratios}.

\begin{figure}
\epsfig{file=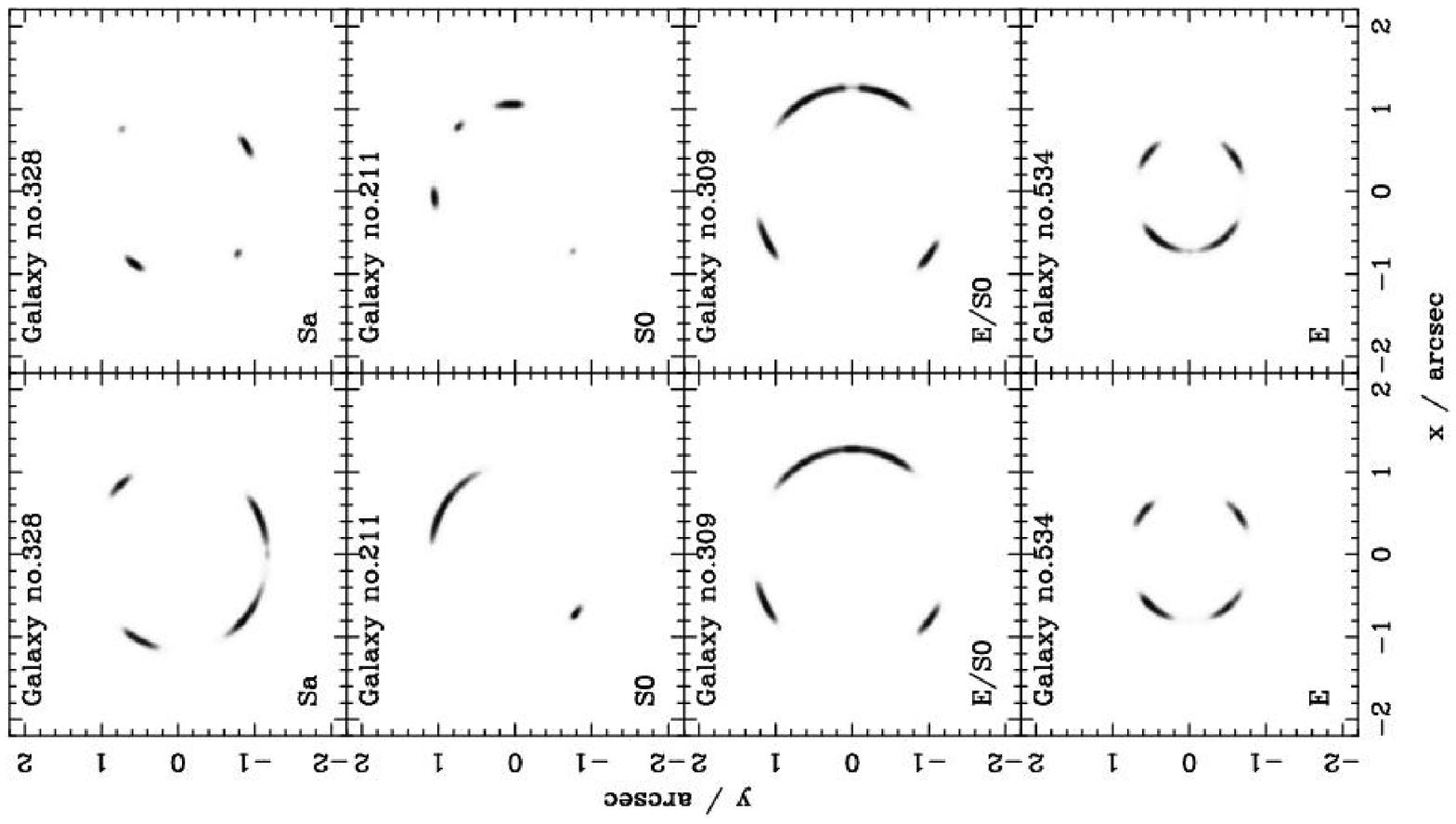,width=14.0cm,angle=-90}\\
\caption{Lensing of an extended source by galaxies with and without
discs. The grey-scale shows the images of a small extended source of
size $\sim0.1\arcseca$ at $z_{\mathrm{s}}=1$ that lies behind lens
galaxies with the properties of the labelled galaxies in
Table\,\ref{table.all}. The models in the right panels include a disc
at an inclination angle of $i=75\,\mathrm{deg}$: the bulge and disc
components are oriented at $45\deg$ with respect to the x-axis in all
cases. The left panels show the corresponding images for bulge-only models.}
\label{properties.ind.fig}
\end{figure}

The importance of discs can also be seen when the relative shifts of the image
positions for a source at a fixed position is considered.  The shift gives some
indication of how significant a given modification of a lens model is.  We show
the image geometries for small, extended sources for a number of our lens models
in Fig.\,\ref{properties.ind.fig}.  There is a small, but measurable change in
the image shapes and positions for E/S0 lenses of the order of $0.05\arcseca$.
In the case of later-type lenses, there is a much more pronounced change; even
Sa lens galaxies significantly change the image geometries.  Note that there is
also a change in the overall scale set by the Einstein radius, of $5-10$\,per
cent.  This is consistent with the change of mass projected within the 
Einstein radius
shown in Fig.\,\ref{fig.mass}.  However, this change of scale only affects the
mass normalization and does not reflect an observable effect of the disc
component.

\subsection{Statistical lensing properties}
\label{cross}
First we determine cross sections
for multiple images and  high magnifications  for  ellipticals that
include an inclined, thin, exponential disc.

\begin{figure}
\epsfig{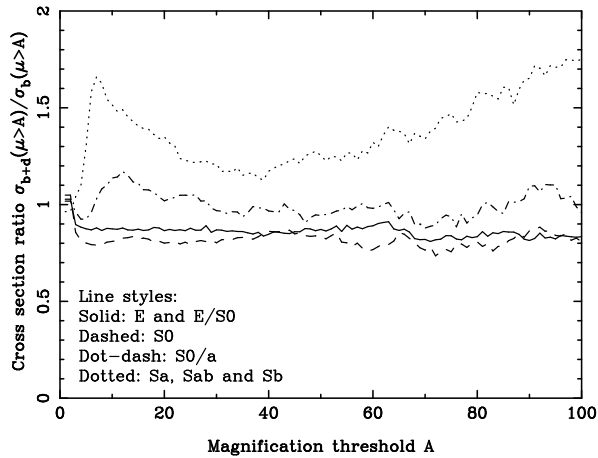}\\
\caption{The ratio of the magnification cross sections.
The curves
show the ratio of the cross sections, $\sigma_{\mathrm{b+d}}(\mu>A)/\sigma_{\mathrm{b}}(\mu>A)$, for total magnifications $\mu>A$ of a
point source at $z_{\mathrm{s}}=1$ as a function of the threshold
magnification, $A$. The different curves are calculated by averaging
over all the galaxies in Table\,\ref{table.all} that have the same
morphological type. When a disc is included, the inclination angle 
$i=75\,\mathrm{deg}$ in all cases.}
\label{properties.cross.fig}
\end{figure}
In  Fig.\,\ref{properties.cross.fig} we  show the  cross section
ratios for bulge and bulge-plus-disc models  as a function  of
magnification, averaged  over  all  the galaxies   of
a   certain morphological  classification   in   the sample. Except for
late-type (Sa, Sab and Sb) galaxies, there is  no
significant increase in the  cross sections  for magnifications  of
up to 100. Even though a  few early-type galaxies show a  clear
signature of an enhancement by the disc component in 
the magnification map on the source
plane, the average effect is small.

\nocite{rusin2001}{Rusin} \& {Tegmark} (2001) showed that the fraction of 4 image systems found in
the CLASS survey is inconsistent with the predictions from standard
lens models. They found that under the assumption of dark matter halos
that are not significantly more flattened than the visible mass, about
$25$\,per cent of all CLASS lenses should be quads. However, the observed
fraction is about $60$\,per cent. We now investigate whether disc structures in
the early-type lensing galaxies could explain this discrepancy.
\begin{figure}
\epsfig{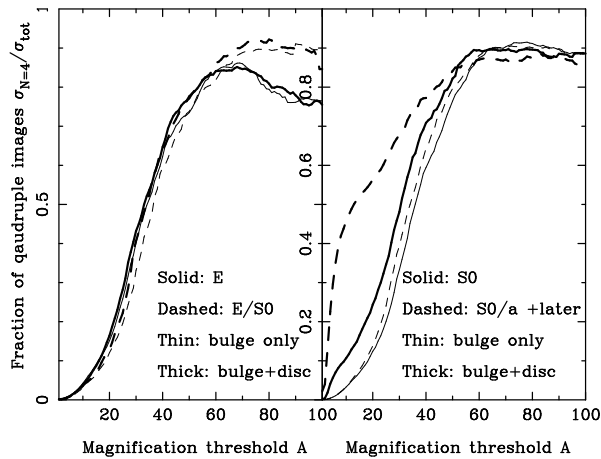}\\
\caption{The fraction of four-image lens systems as a function of
minimum magnification for different lens models. For each
morphological class of galaxies listed in Table\,\ref{table.all} the
average cross section ratio
$\sigma_{\mathrm{N=4}}/\sigma_{\mathrm{tot}}$ is shown, where
$\sigma_{\mathrm{N=4}}$ is the cross section for the formation of 4
images and $\sigma_{\mathrm{tot}}$ is the total cross section for
multiple imaging. The left panel shows the results for earlier types, the right panel for later types. In all bulge-plus-disc models, a disc with the parameters
as listed in Table\,\ref{table.all} is included, with an inclination
angle $i=75\,\mathrm{deg}$.}
\label{properties.nums.fig}
\end{figure}
Fig.\,\ref{properties.nums.fig}   shows  the  expected fraction of
quad systems as a function of the total magnification threshold
$A$ for the different morphological types.  There is no
significant increase in the cross  section for the formation of a quad lens
for the subset of early-type  galaxies; this cross section 
increases significantly only for the late-type
galaxies. Many observed quad lens systems are very symmetrical. This 
suggests that the sources do not lie very close to a cusp
(B1422+231 is an exception), and so that the
sources of the majority of known quad lens systems are likely to be magnified
by only moderate amounts $\mu\sim10-20$.  Hence,
discs in ellipticals cannot explain the
discrepancy between the predicted and observed quad fraction.
However, our calculations are conservative, as they do not take
into account the proper statistical range of lens parameters, source
luminosity function, the selection function of the survey and the
magnification bias. \nocite{rusin2001}{Rusin} \& {Tegmark} (2001) show that such factors can increase 
the predicted fraction of quads at modest magnifications by up to a factor of
$\sim2$. 

Statistical lensing constraints obtained from CLASS  
by \nocite{chae2002}{Chae} (2002) suggest that the observed fraction of 
quad lenses is consistent with an average lens ellipticity $e\sim0.4$, and thus 
that many S0 and later lenses must be present to 
match this ellipticity.
Our results in Fig.\,\ref{properties.nums.fig} suggest that if 
the majority of lens galaxies are indeed later than S0, then 
the predicted fraction of quads is consistent with the observations.

In summary, these results indicate that the  presence of discs with 
constant mass-to-light ratio in ellipticals,
is unlikely to affect the
population statistics of strong lens systems.  However, inspecting
the magnification maps on the source plane shows that discs
might have a significant effect for individual lenses
(Fig.\,\ref{properties.ind.fig}). To investigate this further we
determined how  much the magnification ratios of 
individual lens systems are affected when a disc component is 
included in the
lens model.


\section{Magnification ratios} \label{magratios}

Attempts to model a number of known quad lens systems, like B1422+231 and
PG1115+080 have shown that smooth power-law ellipsoidal models cannot fit both the
observed image positions and magnification ratios.  This has been generally
interpreted as evidence for substructure in their CDM haloes
\nocite{metcalf2001,bradac2002,chiba2002,dalal2002,metcalf2002}({Metcalf} \&
{Madau} 2001; {Brada{\v c}} {et~al.}  2002; {Chiba} 2002; {Dalal} \& {Kochanek}
2002; {Metcalf} \& {Zhao} 2002).  The observational detection of substructure in
dark matter haloes would be of great importance for the standard model of
structure formation, which predicts such CDM substructure on galactic and
sub-galactic scales \nocite{moore1999}({Moore} {et~al.}  1999).  If the presence
of CDM halo substructure provides the only possible explanation for the observed
magnification ratios, this would provide strong observational evidence in support
of the current CDM model.  We now investigate whether discs
in early-type lens galaxies provide an alternative explanation.

\subsection{Causes for unusual magnification ratios}
As pointed out by \nocite{mao1998b}{Mao} \& {Schneider} (1998), for point sources
that are highly magnified and split into 4 magnified images, A, B, C and D, with B
the brightest and D the faintest image (cf.  Fig.\,\ref{properties.sing.fig} in
the previous section; note that we differ here from the more conventional
labelling, where images are labelled A,B,C and D in order of decreasing
brightness), the individual image magnifications obey the relation

\begin{equation}
\mu_{\mathrm{AC/B}}\equiv\frac{\mu_{\mathrm{A}}+\mu_{\mathrm{C}}}{\mu_{\mathrm{B}}}=1,
\label{theorem}
\end{equation}
in the limit of $\mu_{\mathrm{tot}}\rightarrow\infty$. The loci of
source positions at which $\mu_{\mathrm{tot}}$ is infinite form the
caustic line, and so equation\,(\ref{theorem}) holds for sources lying
exactly on the caustic. What is not known, however, is how well
equation\,(\ref{theorem}) holds for sources that lie close to
the caustic, or, for sources of small but finite size. In
addition, it is not clear how the details of the lensing potential
will affect this behaviour close to the caustics.  The observed
magnification ratios in systems like B1422+231 
grossly violate equation\,(\ref{theorem}). This discrepancy 
is the main reason
why the observed magnification ratios are generally regarded as 
strong evidence for CDM-halo substructures with 
masses of the order of $10^6-10^7$\,M$_{\odot}$, which can 
strongly affect the flux of individual images 
if located close to the images.
However, there are several other explanations:
\begin{enumerate}
\item microlensing. In many strong lens systems, microlensing variability 
has been detected and it is difficult to exclude the possibility 
of microlensing
in the observed systems with certainty.
\item differential extinction. There is a possibility that optical
flux ratios of the images are affected by different dust columns along
the lines of sight through the lensing galaxy.
\item the source is extended.  Extended radio sources may lead to very
different magnification ratios in the observed images, depending on
the size and relative brightness of the different source
components. Recent simulations by \nocite{moustakas2003}{Moustakas} \& {Metcalf} (2003)
suggest that even for extended sources there is no change in
the continuum magnification ratios for smooth lens models; only the relative contributions from broad and narrow lines is affected.
\item off-caustic sources. If the source does not lie
exactly on the caustic but is still magnified strongly, then the
magnification ratios will not necessarily obey equation\,(\ref{theorem}).
\end{enumerate}
We now investigate the 
validity of the last explanation in the context of discs in
early-type lensing galaxies.

\subsection{Known lens systems with unusual magnification ratios}
There have been two recent studies of lens systems with unusual magnification
ratios.  In the comprehensive study by \nocite{metcalf2002}{Metcalf} \& {Zhao}
(2002) 4 systems are listed where the magnification ratios deviate by
$\sim0.1\,\mathrm{mag}$ or more from the expected values assuming a one
component smooth ellipsoidal lens with external shear; B1422+231, PG1115+080,
H1413+117 and Q2237+030.  The first three systems are quasars lensed by
early-type galaxies while the lens in Q2237+030 is a barred spiral.  Of the four
systems, H1413+117 lies close to a cluster of galaxies which may crucially
affect the lensing potential \nocite{kneib1998,chae1999}({Kneib} {et~al.}  1998;
{Chae} \& {Turnshek} 1999).  Q2237+030 is lensed by a galaxy at a very low
redshift of $z=0.04$ and microlensing affects the flux ratios by up to a factor
of two \nocite{irwin1989,wozniak2000}({Irwin} {et~al.}  1989; {Wo{\' z}niak}
{et~al.}  2000).  PG1115+080 and B1422+231 are affected only very moderately if
at all by microlensing \nocite{vanderriest1996,yee1996}({Vanderriest} {et~al.}
1986; {Yee} \& {Bechtold} 1996), but both reside very close to, or even within,
a galaxy group \nocite{tonry1998,kundic1997a,kundic1997b}({Tonry} 1998;
{Kundi{\'c}} {et~al.}  1997a,b).  As shown by
\nocite{moller2002}{M{\"o}ller} {et~al.}  (2002), the presence of the group can
significantly alter the lensing behaviour and it is doubtful whether a simple
power-law-plus-shear model, as used for example by \nocite{keeton1997}{Keeton},
{Kochanek} \& {Seljak} (1997) reflects the true lensing potential adequately.
For B1422+231 the predicted and observed magnification ratios differ
by up to $0.3\,\mathrm{mag}$, as compared to only 
$0.1\,\mathrm{mag}$ for PG1115+080,
and so it is less likely that the discrepancy in B1422+231 is a result of the
details of the group mass distribution surrounding the lens.  This
points out B1422+231 as the strongest candidate for dark
substructures:  it is studied in great detail by \nocite{bradac2002}{Brada{\v
c}} {et~al.}  (2002).  

A recent theoretical study 
\nocite{keeton2002}({Keeton}, {Gaudi} \& {Petters} 2002)
investigated the
magnification ratios of four caustic lens systems in more detail, 
using an analysis similar to
that presented here but with a simple one-component lens model:
B2045+265, B0712+472, RX J0911+0551 and
B1422+231 \nocite{keeton2002}({Keeton}, {Gaudi} \& {Petters} 2002).  RX
J0911+0551 has only optical and near-infrared flux measurements, and so the
extinction corrections for this source are highly uncertain.
B0712+472 is possibly affected by a foreground group
\nocite{fassnacht2002}({Fassnacht} \& {Lubin} 2002).  Keck spectra of B2045+265
indicate that it is likely to be a late-type lens
\nocite{fassnacht1999}({Fassnacht} {et~al.}  1999), and so it is unlikely that a
simple one-component model is viable:  a disc is almost
certainly present.  Note that only B1422+231 was included in the study
of \nocite{metcalf2002}{Metcalf} \& {Zhao} (2002).  

The number of lens
systems that are claimed to have magnification ratios that can only be explained
using compact CDM substructure is rather large.  In some cases,
microlensing variability or differential extinction could be a viable
alternative explanation, but not for all.

\begin{table*}
\caption{
Summary of observed quad systems with unusual magnification ratios.
$^1$ {Christian}, {Crabtree} \&  {Waddell} 1987. The fluxes are normalized to image D.
$^2$ {Jackson} {et~al.} 1998. The listed values are for the Merlin 1996 5 GHz data.
$^3$ {Fassnacht} {et~al.} 1999. The listed values correspond to the VLA 8.5GHz data.
$^4${Patnaik} {et~al.} 1999. The listed values correspond to the VLA 8.5GHz data.
$^5$ {Chae} \& {Turnshek} 1999. All images have been relabelled to correspond to the convention adopted in other systems.
$^6$ {Falco} {et~al.} 1996. All images have been relabelled to correspond to the convention adopted in other systems.
$^7$ {Burud} {et~al.} 1998. The fluxes are normalized to image A.
}
\begin{tabular}{l|r|r|r|r|r|r}\hline
System & $F_{\mathrm{A}}$ & $F_{\mathrm{B}}$ & $F_{\mathrm{C}}$ & $F_{\mathrm{D}}$ & $\mu_{\mathrm{AC/B}}$ & $\mu_{\mathrm{ABC/D}}$\\ \hline
PG1115+080$^1$ & 2.49$\pm$0.03 & 3.22$\pm0.03$ & 0.64$\pm0.03$ & 1.0 & 0.98$\pm0.1$ & 6.35$\pm0.1$\\
B0712+472$^2$ & 10.5$\pm0.1$mJy & 8.5$\pm0.1$mJy & 4.7$\pm0.1$mJy & 0.9$\pm0.1$mJy & 1.78$\pm0.03$ & 26.3$\pm2.1$ \\
B2045+265$^3$ & 18.4$\pm0.3$mJy & 9.42$\pm0.3$mJy & 14.8$\pm0.3$mJy & 2.41$\pm0.3$ mJy & 3.5$\pm0.1$ & 17.7$\pm2.1$ \\
B1422+231$^4$ & 152$\pm2$ & 164$\pm2$ & 81$\pm1$ & 5$\pm0.5$ & 1.42$\pm0.03$ & 79.4$\pm7.9$ \\
H1413+117$^5$ & 1.65$\pm0.16$ & 1.62$\pm0.16$ & 1.0 & 0.67$\pm0.1$ & 1.63$\pm0.16$ & 6.1$\pm0.6$\\
Q2237+030$^6$ & 78$\pm14$ $\mu$Jy & 60$\pm14$ $\mu$Jy & 85$\pm14$ $\mu$Jy & 43$\pm14$ $\mu$Jy & 2.7$\pm0.8$ & 5.2$\pm2.4$ \\
RX J0911+0551$^7$ & 1.0 & 0.965$\pm0.013$ & 0.544$\pm0.025$ & 0.458$\pm0.004$ & 1.60$\pm0.08$ & 5.48$\pm0.28$  \\
\hline
\end{tabular}
\label{table.systems}
\end{table*}

\nocite{christian1987}
\nocite{jackson1998}
\nocite{fassnacht1999}
\nocite{patnaik1999}
\nocite{chae1999}
\nocite{falco1996}
\nocite{burud1998}
We summarise the properties of all these lens systems in Table\,\ref{table.systems}.
Together with the image fluxes, we list the magnification ratios 
$\mu_{\mathrm{AC/B}}$ and 
$\mu_{\mathrm{ABC/D}}\equiv 
(\mu_{\mathrm{A}}+\mu_{\mathrm{B}}+\mu_{\mathrm{C}})/\mu_{\mathrm{D}}$.

\subsection{Magnification ratios for bulge-plus-disc models of early-type lens galaxies}

To answer the question whether discs in early-type lenses could explain 
some or all of the
observed magnification ratios, we computed all the image multiplicities, 
position and magnifications 
for all sources on a $400\times400$ pixel grid for 
our particular representative model galaxy (galaxy no. 242).
We performed the calculations both on a low-resolution grid that contained
the whole caustic structure and on a smaller, higher resolution grid, that 
contained one of the cusps along the major axis of the lens galaxy. For both 
grids, we selected all sources with quad images, yielding a total sample of 
about 
40000 simulated image systems.

\begin{figure}
\epsfig{file=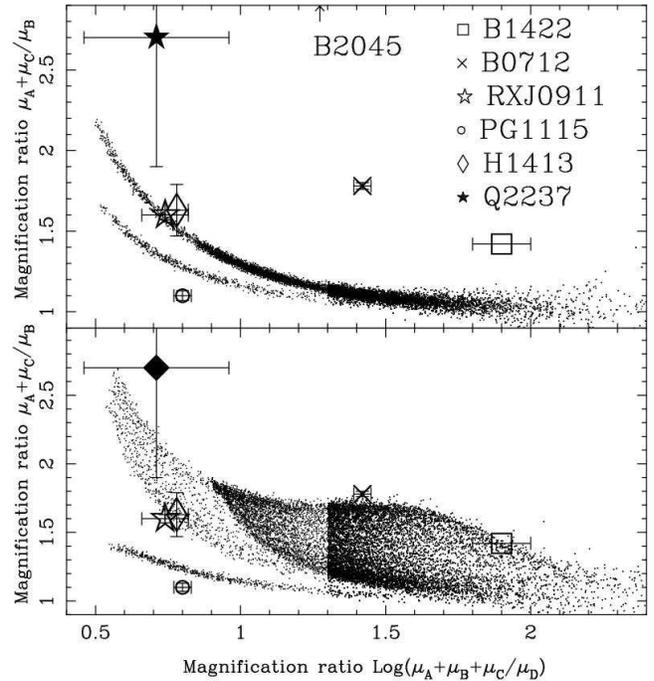,width=9.0cm,angle=-90}\\
\caption{Scatter diagrams of magnification ratios for quad images
of point sources lensed by a model galaxy with parameters matching
galaxy no.\,242 in Table 1. The markers in the upper panel show the
magnification ratios between images produced by the bulge-only model,
whereas the markers in the bottom panel show the results for the
bulge-plus-disc model. For total magnifications below 20 only 10\,per cent 
of all simulated image systems are shown. Note that the images in the
bulge-plus-disc case fall into two regions, whereas all the images produced
by pure-bulge models obey equation\,(\ref{theorem}) in the limit of
$\mu_{\mathrm{ABC/D}}\rightarrow 1$. The
large markers indicate the magnification ratios for the observed systems with the $1\sigma$
error, as listed in Table\,2. 
A small scatter of about 
10\,per cent is introduced at high $\mu$ due to numerical uncertainties.
}
\label{magrat.scatter.fig}
\end{figure}

Fig.\,\ref{magrat.scatter.fig} shows a scatter plot of the ratio of
the sum of the magnifications of the outer two images to the central
image $\mu_{\mathrm{AC/B}}$, versus the ratio of the total magnification of the brightest
three images, A-C to the magnification of the faintest image D, $\mu_{\mathrm{ABC/D}}$.
The points in the upper panel are for the pure-bulge model, whereas
the points in the lower panel are for the bulge-plus-disc model. For large
magnification ratios there is a significant difference.
All the images of sources with high total magnifications,
corresponding to a high
$\mu_{\mathrm{ABC/D}}$
ratio, scatter around
$\mu_{\mathrm{AC/B}}\sim1$ for the
bulge-only models. For large $\mu$ the  scatter of a few per cent 
equals that expected from numerical uncertainties in the
simulations, due to 
 small deviations in the deflection angle that 
lead to large deviations in the source position and images occasionally 
being missed or 
associated with a wrong source position.
Note that there are two distinct and well defined `populations' of images, corresponding to
images in the two different types of cusp regions (see  
\begin{figure}
\epsfig{file=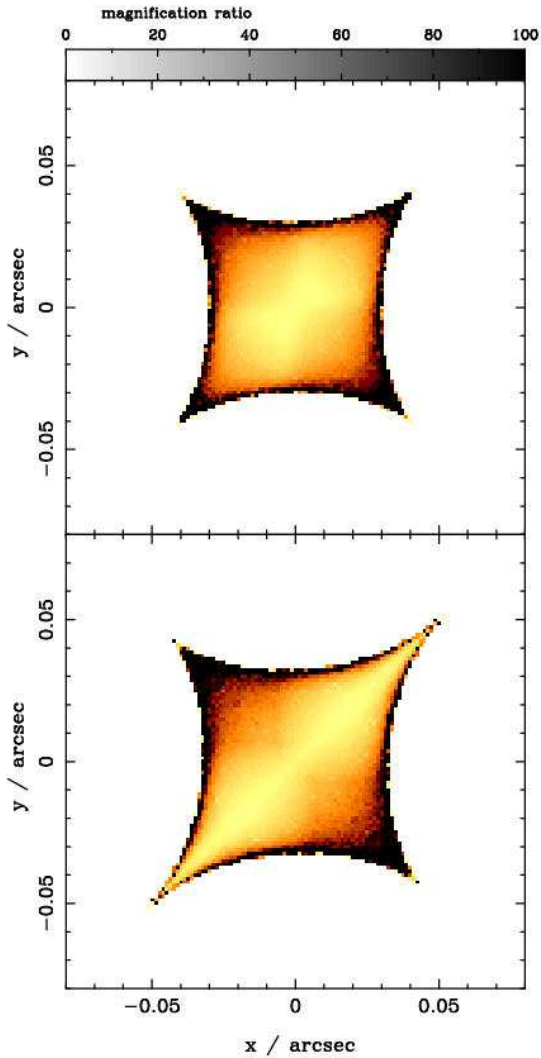,width=14.0cm,angle=-90}\\
\caption{Maps of the magnification ratio
$\left[(\mu_{\mathrm{A}}+\mu_{\mathrm{C}})/\mu_{\mathrm{B}}-1\right]^{-1}$
as a function of source position. The upper panel is for the
bulge-only model, the lower panel for the bulge-plus-disc model. Large
values (dark shades) indicate that the magnification ratios of the
images of a source at that position obey
equation\,(\ref{theorem}). Small values (light shades) indicate that the
image magnification ratios deviate from that relation. Note that
Fig.\,\ref{properties.sing.fig} shows an example configuration for
sources lying in the bottom left cusp.}
\label{magrat.maps.fig}
\end{figure}
Fig.\,\ref{magrat.maps.fig}). The two panels of Fig.\,\ref{magrat.maps.fig} show quantitatively how much the image
magnification ratios deviate from equation\,(\ref{theorem}) as a
function of source position, both for the bulge-plus-disc model (upper
panel) and the bulge-only model (lower panel). These maps are similar
overall, but differ significantly in the two cusps that are aligned with the
disc: when a disc is present, images of sources that lie
very close to these cusps have magnification ratios that are
inconsistent with the predictions of equation\,(\ref{theorem}). The
images that are produced by sources that lie close to these two cusps
form the second population
of images in the bottom panel of Fig.\,\ref{magrat.scatter.fig}. 

As an example, Fig.\,\ref{properties.sing.fig} illustrates the behaviour for sources
in the lower-left cusp of our model. By comparing the relative fluxes of the
images in the top and bottom panels of Fig.\,\ref{properties.sing.fig}, 
one can see by eye that the sum of the fluxes of
the images A and C is roughly equal to the flux of image B in the bulge 
only model, but not in the bulge-plus-disc model. The relative positions 
and flux ratios of
the images A,B and C in the lower panel of that figure are very close
to those observed in B1422+231.

Figs\,\ref{magrat.scatter.fig} and \ref{magrat.maps.fig} show the difference between the bulge and bulge-plus-disc models. In 
the bulge-plus-disc model a set of
sources exists with an image magnification ratio $\mu_{\mathrm{AC/B}}$ 
very different from one,
even for high magnifications of $\mu_{\mathrm{ABC/D}}\sim50$. All these sources lie inside one of the two cusps along the major axis.
Note that for all those images, the faintest image D
lies at a large distance of $\sim1.2\arcseca$ from the centre of the
lensing galaxy and is not demagnified. This reflects the high total
magnification of a source which lies close to the caustic line.

Some unusual observed magnification ratios of some observed lens systems are 
indicated by the large symbols in Fig.\,\ref{magrat.scatter.fig}.
Error bars indicate the 1-$\sigma$ flux measurement errors (Table\,\ref{table.systems}).
Only H1413+117 and RXJ0911+0551 lie in regions of the 
diagram that are consistent with a bulge-only model. 

However, the 
bulge-plus-disc model is clearly consistent with the 
magnification ratios for
PG1115+080, H1413+117, RX J0911+0551, B0712+472, Q2237+030 and B1422+231. 
B2045+265 is the only system for which our model cannot reproduce 
the observed magnification
ratios. This source is lensed by a spiral galaxy, and so
our early-type lens model with a very modest
disc would not be expected to provide a good fit. 

\begin{figure}
\epsfig{file=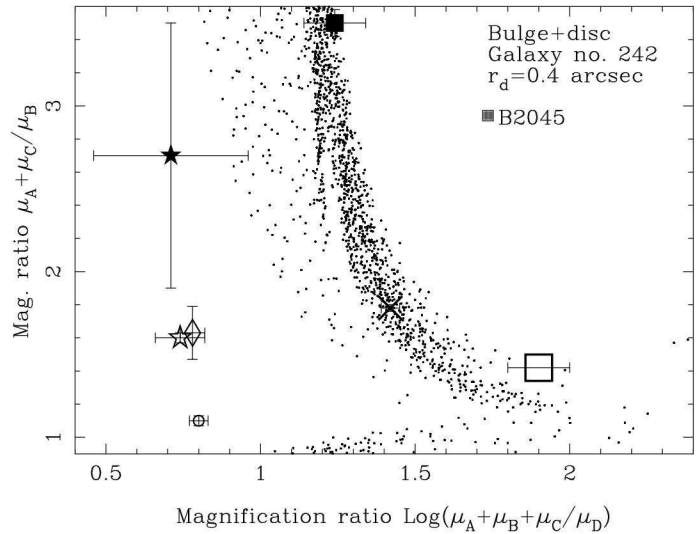,width=7.0cm,angle=-90}\\
\caption{Scatter diagrams of magnification ratios for quad images
of point sources lensed by a model galaxy with parameters as in the Fig.\,9, 
but with
a heavier disc that has twice the size. The symbols are in the previous plot.
}
\label{magrat.scatterdisc.fig}
\end{figure}
In Fig.\,\ref{magrat.scatterdisc.fig} we show magnification ratios for about 
40000 simulated image systems (without using a second higher-resolution grid) 
for a bulge-plus-disc model with a disc that has a scale-length twice 
that of the previous bulge-plus-disc model. All the other parameters are 
unchanged and correspond to galaxy no. 242. The effect of increasing 
the disc mass is to substantially increase the $\mu_{\mathrm{AC/B}}$ ratio for 
image systems with magnifications $\mu_{\mathrm{ABC/D}}\sim20$. The 
magnification ratio of B2045+265 is now consistent with the predicted value.

We conclude that the presence of an inclined disc component containing
of order 
5\,per cent of the total mass can describe all of the known early-type lens
systems with unusual magnification ratios without requiring compact substructure
in the lens halo.

\subsection{Effect of disc size, disc inclination, bulge ellipticity and a dark halo}
\label{propdep}

We have used a particular model galaxy (\S\,\ref{surbripro}) as a lens.  In
order to make sure that our results apply in general to bulge-plus-disc models
of early-type galaxies and are not specific to our particular choice of
parameters we investigated a range of other models.  The full 
parameter space for
bulge-plus-disc models is large and a detailed study is beyond the scope of this
paper.  We investigated the dependence on several parameters
qualitatively.  We find that, qualitatively, any disc that \emph{on its own}
produces a caustic of similar or larger size than that produced by the bulge is effective at modifying the image properties.
In practice that means any disc that is inclined by more than $60\,\deg$ with a
scale-length of the order of the size of the caustic of the corresponding
bulge-only model.  For nearly spherical bulges, discs with smaller 
scale-lengths are also 
effective, and naturally produce magnification ratios similar to those
observed for B1422+231.  For disc sizes below $0.2\,\arcseca$, the disc
inclination must be increased beyond $i=80\,\deg$ to obtain magnification ratios
consistent with B1422+231.  Even an edge-on disc smaller than $0.1\arcseca$
cannot reproduce the magnification ratios of B1422+231.
There is also the possibility of misalignment between the positions of the disc and the bulge/halo. Such misalignments are to be expected as a result of recent merger events and may lead to complicated and unusual lensing configurations 
and provide more extreme changes in the magnification ratios, as discussed by 
\nocite{quadri2002}{Quadri}, {M\"oller} \&  {Natarajan} (2002).

We have also
investigated models with a CDM halo.  Adding a singular CDM halo, containing 2/3
of the total mass within an Einstein radius, and adjusting the mass
normalization of the bulge and disc so as to keep the Einstein radius constant,
does not significantly alter the effect of the disc.  It is purely the presence
of a thin disc structure in combination with a 
more spherical mass component 
that produces the change in magnification ratios.

\subsection{Expected number of affected systems}
\nocite{moller1998}{M{\"o}ller} \& {Blain} (1998) have shown that discs in spiral galaxies are
important for the statistical effects on lensing cross sections 
only at inclination angles of $\ge
60\,\deg$. Assuming that a fraction of $50-100$\,per cent of all early-type
lens galaxies do contain discs, $15-35$\,per cent will contain discs with
inclination angles of $\ge 60\,\deg$. A conservative estimate of the
fraction of early-type lens systems in which discs may be important is
thus $\sim10-30$\,per cent. This fraction is consistent with the number of
CLASS quad lens systems with deviant magnification ratios:
of the 7 CLASS quad lenses, at least two early-type lens system
(B1422+231 and B0712+472) exhibit magnification ratios
inconsistent with simple power-law plus external-shear models.

An
additional factor to consider is magnification bias. Depending on the
bulge-to-disc ratio, the disc may lead to higher cross sections
for strong magnifications, as shown in
Fig.\,\ref{properties.cross.fig}. This could bias a flux-limited
sample towards lens galaxies that contain discs. However, as shown in
Fig.\,\ref{properties.cross.fig}, the cross section is not increased
significantly for E/S0/Sa galaxies, and so the 
magnification bias is likely to be small.

\subsection{Distinguishing the effect of discs and CDM substructure}

The results described in this section demonstrate that discs in early-type 
lensing galaxies
may explain some of the observed magnification ratios that have
hitherto only been explained assuming CDM substructure.
There are some key observational differences between these two scenarios.

First, discs in early-type lens galaxies are, in principle,
detectable by bulge-plus-disc light profile decompositions as discussed in
\S\,\ref{surbripro}. However, the light of the background quasar
makes accurate photometry of lens galaxies extremely difficult in
practice. \nocite{treu2002}{Treu} \& {Koopmans} (2002)  obtained kinematic
information of a lensing galaxy from spectroscopic observations, and  
demonstrated that it is thus 
possible to obtain additional constraints on its mass
distribution. It might be possible to 
constrain possible disc structures in lensing galaxies in this
way. 

Secondly, disc structures significantly affect only the 
magnification ratios of
sources that lie within one of the two cusps along the discs. Assuming
that discs are oriented along the major axis of the bulge ellipticity, this
means that disc structures provide an explanation of discrepant
magnification ratios only for quad image systems with 
three strongly magnified images along a line perpendicular to the
major axis of the lens galaxy, whereas CDM substructure could modify 
any image configuration. Note that this is \emph{not} true 
for late-type lenses with extended inclined discs, in which 
the caustic structures are completely dominated by the disc, and the 
disc strongly affects the magnification ratios of quad systems produced 
by sources in any one of the four cusps.
 
We stress that these results do \emph{not} prove the existence of discs in early-type lensing galaxies, nor do we propose that such disc structures provide the only explanation for all the observed magnification ratios. 
For all systems, the observed magnification ratios could be caused by CDM 
substructures. We have not carried out detailed modelling of any of the 
observed systems, and our analysis does not include joint fitting to both 
magnification ratios and image positions. However, in contrast to 
\nocite{keeton2002}{Keeton} {et~al.} (2002) we find that magnification ratios alone do not provide 
conclusive evidence for the presence of CDM halo substructure: 
realistic disc structures can explain the observed magnification ratios 
equally well. Detailed case-by-case lens modelling using bulge-plus-disc models is now needed to verify that early-type galaxies with disc structures are indeed viable models for all or at least some of the observed lens systems. The fact that discs affect image positions very slightly, even for a fixed source position, indicate that detailed bulge-plus-disc models are likely to be successful 
in fitting magnification ratios and image positions.

\section{Effect on Hubble parameter estimation}
\label{hubble}
Strong gravitational lensing of variable background quasars provides a unique
method to determine the Hubble constant. Brightness fluctuations 
are observed at different times
at each image as the path lengths and gravitational time
delays to each image depend on its relative position and the local lensing
potential. If the image positions and lensing potential are
known, then $H_0$ can be determined, as $\Delta
t\propto1/H_0$. Gravitational and geometric time delays are often of
roughly equal magnitude, and so an accurate knowledge of the lensing
potential is crucial to deduce an accurate value of $H_0$.

Lens systems with measured time delays are nearly always associated
with early-type lens galaxies, although 
B1600+434 \nocite{koopmans2000}({Koopmans} {et~al.} 2000)
and B0218+357 (Wucknitz, priv. comm.) are 
notable  exceptions. These systems have been modelled using
bulge-only models, with a constant external shear in
some cases.

We now investigate how a disc component is likely to
affect time delays for the bulge and bulge-plus-disc model of galaxy
no.\,242. The time delay between two images at
$\vec{\theta}_1$ and $\vec{\theta}_2$ of a source at $\vec{\beta}$
is
\begin{equation}
\Delta
t=\frac{D_{\mathrm{OS}}D_{\mathrm{OL}}}{cD_{\mathrm{LS}}}(1+z_{\mathrm{l}})\left[\frac{(\vec{\beta}-\vec{\theta_1})^2-(\vec{\beta}-\vec{\theta_2})^2}{2}-\Delta\psi_{\mathrm{b}}+\Delta\psi_{\mathrm{d}}\right],
\label{timedelay.eq}
\end{equation}
where $D_{\mathrm{OL}}$ is the angular diameter distances between
observer and lens.  The distances and thus $\Delta t$ are
proportional to $H_0^{-1}$. The lens redshift is
$z_{\mathrm{l}}$ and the potentials due to bulge and disc components
are $\psi_{\mathrm{b}}$ and $\psi_{\mathrm{d}}$ respectively.  
In \S\,\ref{properties} we pointed out that the shift in image positions between bulge-plus-disc and bulge-only models for E and E/S0 type lensing galaxies is expected to be very small. 

We are concerned here with the effects of modelling an early-type lens that contains a disc with a bulge-only model. Since the lens models are obtained from fits to the image positions, the image positions remain the same irrespective of the model used.  
Therefore, the geometrical time delay of a given pair of images is not affected by a disc component; any change in the time delay is due to the 
change in the potential at the image positions.

Note that compact CDM substructure does not affect time-delays measurably.

\subsection{Four-image/quad systems}

About half of the known lens systems with measured time delays are quad systems.
Since the quad caustic structure is the region in the source plane that is most
strongly affected by the presence of a disc, the greatest effect of a disc on
time delays is expected in these systems.  In most cases a single time delay is
measured, usually between the image pair with the largest separation.
Using our ray-tracing code in combination with equation\,(\ref{timedelay.eq}) we
calculated the time delays between images B and D for all quad systems of the
model galaxy described in \S\,\ref{surbripro}.
\begin{figure}
\epsfig{file=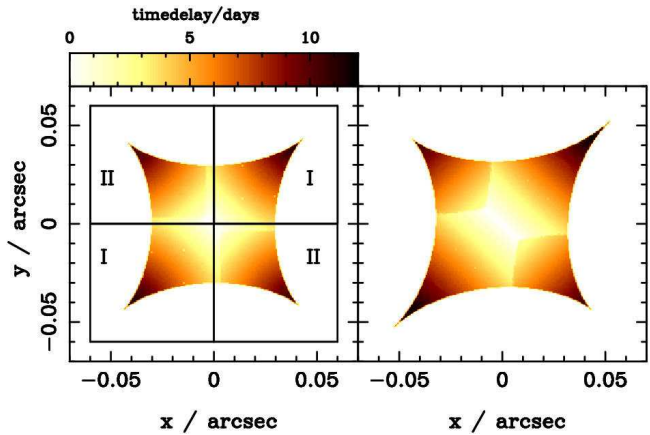,width=5.7cm,angle=-90}\\
\caption{Time delay as function of source position. The two maps show the time delays between images B and D (cf. Fig.\,\ref{properties.sing.fig}) for all quad geometries for model galaxy no.242. The left panel shows the results for the bulge-only model and the right panel the results for the bulge-plus-disc model. The boxes in the left panel mark the different cusp regions, as discussed in the text.    
}
\label{timedelay.map.fig}
\end{figure}
Fig.\,\ref{timedelay.map.fig} shows the time delays as a function of source position for bulge-plus-disc and bulge-only models. The difference is on the order of a few tens of per cent, and is largest in the cusps lying along the disc (the regions marked I). Especially in the cusp regions close to the caustic, where the 
image magnification is greatest, the time delays for the bulge-plus-disc 
model are $20-30$\,per cent larger than for the bulge-only model.

\begin{figure}
\epsfig{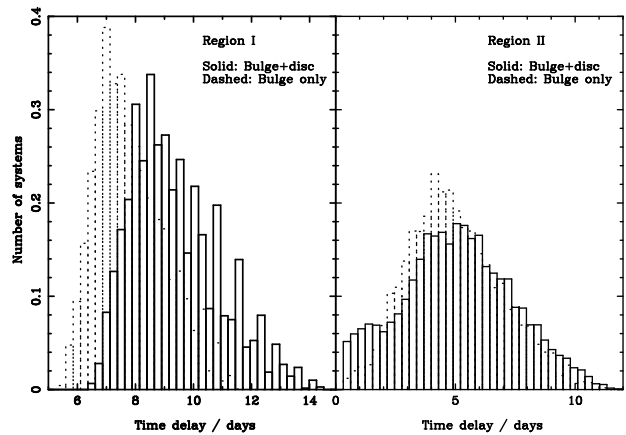}\\
\caption{Histogram of time delays. This plot shows the
expected relative fraction of systems as a function of timedelay between images B and D. The left panel shows the statistics for the regions marked I in Fig.\,\ref{timedelay.map.fig} and the right panel the statistics for the regions marked II. The histograms are shown for our bulge-plus-disc and bulge-only lens galaxy model (see \S\,\ref{surbripro}).
Note that the time delay is smaller for bulge-only systems, due to the shallower potential.}
\label{timedelay.td.fig}
\end{figure}
The time-delay statistics for regions I and II are shown in Fig.\,\ref{timedelay.td.fig}. For each region in 
Fig.\,\ref{timedelay.map.fig} the time delays are divided into 50 bins 
and the histograms are normalized to unity.
Comparing the histograms in the right panel of Fig.\,\ref{timedelay.td.fig},  
there is no significant difference between the bulge-plus-disc and bulge-only model time-delay statistics in the cusps away from the disc (the regions marked II on Fig.\,\ref{timedelay.map.fig}). However, the histograms in the left panel of Fig.\,\ref{timedelay.td.fig} show that the time delays in the cusps lying along the disc are increased by $\sim20$\,per cent for the bulge-plus-disc model with respect to the bulge-only model. The overall distribution for the bulge
only model is shifted towards shorter time delays with respect to the
bulge-plus-disc models. Assuming that an elliptical lens galaxy that
contains a disc is modelled using a bulge-only model, this suggests
that the Hubble constant
will be underestimated systematically. 
\begin{figure}
\epsfig{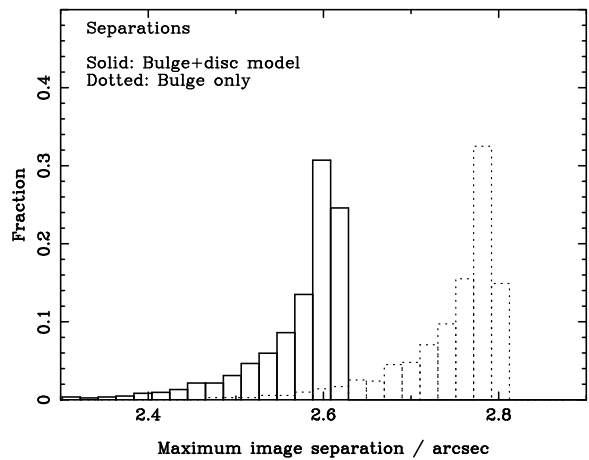}\\
\caption{Histogram of the separations of images. This plot shows the
expected relative fraction of systems as a function of
separation. Note that the separation is larger for bulge-only systems,
due to the higher mass in the central region.}
\label{timedelay.sep.fig}
\end{figure}

In Fig.\,\ref{timedelay.sep.fig} we show explicitly that disc components 
alter the individual image positions only very slightly
for our model galaxy no.\,242.
The figure shows the distribution of image
separations for both models. The images are systematically further
apart in the bulge-only models, suggesting a larger mass inside the
Einstein radius and thus a larger predicted time delay, if the 
radial profile of the potential is unaffected.  In observed
lens systems, the lens would be modelled to fit the image
configurations, and so a bulge-only
model fitted to the observed image separations would contain less
mass within the Einstein radius than the bulge-only model with the
time delay distribution shown in Fig.\,\ref{timedelay.td.fig}. This
would shift the time-delay distributions shown in
Fig.\,\ref{timedelay.td.fig} for the bulge-only model to even lower
time delays. The only possibility both to fit the image separations and
obtain similar distribution of time delays is by changing the gradient of the
potential between the images. For point sources, the only
other lensing property that is sensitive to the shape of the potential
is the magnification ratio between the images. Thus, if 
magnification ratios are not used to constrain lens models, then 
bulge-only models might fit the image positions of a lens that
actually consists of a bulge-plus-disc lens, yielding an incorrect mass
model and an $H_0$ value that is systematically too low by up
to 25\,per cent. 

\subsection{Two-image systems}
Quad lens systems 
are generally regarded as being more useful than 2-image
systems for $H_0$ determinations because more constraints on
the lensing potential are available. However, many 2-image 
lens systems have been used for time delay measurements.
In these cases, the source lies outside 
the diamond-shaped caustic, where magnifications 
$\mu\sim2-10$. These images 
are expected to be affected only weakly by a disc component. 

\begin{figure}
\epsfig{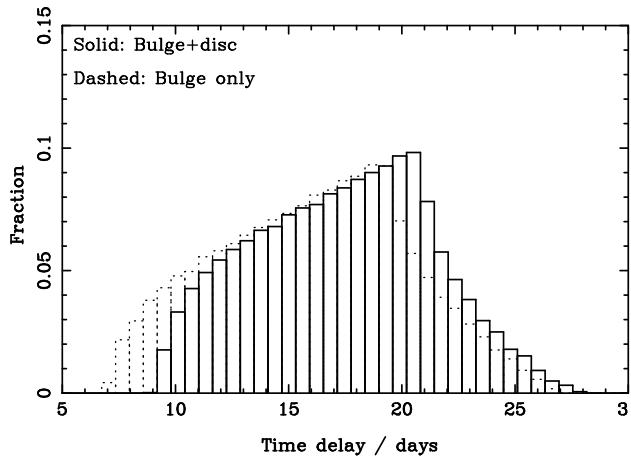}\\
\caption{Histogram of time delays for 2-image systems. This plot shows the
expected relative fraction of systems as a function of time-delay between the two images in 2-image lens systems. The histograms are shown for our bulge-plus-disc and bulge-only lens galaxy model as described in \S\,\ref{surbripro}.}
\label{timedelay.two.fig}
\end{figure}

We have calculated the time delays for 2-image systems and show the corresponding
statistics in Fig.\,\ref{timedelay.two.fig}.  Comparison with
Fig.\,\ref{timedelay.td.fig} shows that the effect of a disc component on the time
delay is much smaller for 2-image systems than for quad systems.  A number of
2-image systems with measured time delays, like B0218+357
\nocite{biggs1999}({Biggs} {et~al.} 1999) have possible disc
structures, but even if these are present, our results indicate that the the
effect on estimates of $H_0$ is small. Note
that recent observations (Wucknitz, private comm.)  indicate that the lens in 
B0218+357 is a spiral galaxy.  Massive discs, as expected in spiral
lenses, may affect the time delays much more strongly. 

\subsection{Parameter dependence} 
The results in
this and the preceding section show that any inclined discs within elliptical lens
galaxies affect sources in the cusps lying along the disc most strongly, whereas
the effect on the cusps lying away from the disc is small.  We have not checked
the dependence of the time-delay statistics on the disc parameters in detail, but
the dependence is similar to that discussed in \S\,\ref{propdep} since
magnification ratios and time delays are both sensitive on changes in the cusps.
The results we obtained in that section also hold for the time delays.  Disc
structures are important if they are inclined at $i>60\deg$ to the line of sight
and are massive/extended enough to produce a caustic of a size that is of the
order of the caustic produced by the bulge.  That is typically the case for discs
larger than $r_{\mathrm{hd}}>0.6\mathrm{kpc}$ and with masses of at least
$\sim5$\,per cent of the bulge, for which time delays will be increased by a few
tens of per cent relative to bulge-only models.  

\subsection{The effect on estimates on $H_0$} 
Our approach here only gives a rough indication of how
determinations of $H_0$ are affected by disc structures.  To be any more
quantitative about the uncertainties that arise it is necessary to perform
detailed modelling of known systems using bulge and bulge-plus-disc models, which
is difficult and time-consuming.  However, our results offer a possible resolution
as to why there is a significant discrepancy between determinations of $H_0$ from
CMB data \nocite{lewis2002}({Lewis} \& {Bridle} 2002), the \emph{HST} Key-project
\nocite{freedman2001}({Freedman} {et~al.}  2001) and gravitational lensing
\nocite{kochanek2003}({Kochanek} 2003).

The effect of disc structures on time-delays is much more significant for quad systems than for 2-image systems. In addition, we expect that there will be a correlation between low-$H_0$ results and unusual magnification ratios if discs are present. 
This seems to be the case: \nocite{biggs1999}{Biggs} {et~al.} (1999) derive 
$H_0=69^{+13}_{-19}\,\mathrm{km\,s^{-1}\,Mpc^{-1}}$ for the 2-image system 
B0218+357, whereas \nocite{impey1998}{Impey} {et~al.} (1998) obtain 
$H_0 = 44\pm4\,\mathrm{km\,s^{-1}\,Mpc^{-1}}$ for PG1115+080, a quad system 
with unusual magnification ratios. The errors on the $H_0$ estimate by \nocite{biggs1999}{Biggs} {et~al.} (1999) have recently been disputed by \nocite{lehar2000}{Lehar}{ et~al.}(2000) and Wucknitz, Biggs \& Browne (MNRAS, submitted).
More recent single-component models of PG1115+080 by 
\nocite{treu2002}{Treu} \& {Koopmans} (2002) 
give a higher value of $H_0=59^{+12}_{-7}\pm 3\,\mathrm{km\,s^{-1}\,Mpc^{-1}}$.
The larger errors are statistical, and the smaller error of $\pm 3$ is an 
estimate of the systematic error.
No other quad systems with measured time delays have unusual magnification ratios, and the values of $H_0$ determined from these systems are generally higher. 

\section{Conclusions}
\label{conclusions}
Gravitational lensing plays a role of ever increasing importance in
determining cosmological parameters, notably $H_0$, and constraining
the dark matter distribution on a variety of scales. More than 50
strong gravitational lens systems are known. 
Generally, it is
not possible to constrain the mass distribution of the lensing galaxy
uniquely, due to the degeneracies present. Whether this
uncertainty in models has serious implications for cosmological
parameter estimation from strong lensing is still unclear.

The majority of all lensing galaxies have been classified as massive
early-type galaxies. Reliable photometric determination of the presence 
of a disc
component is
difficult due to the usually very bright lensed quasar images. 
Therefore, discs similar to those discussed
in this paper would not generally have been identified in existing
images. However, observations
of early-type galaxies in the local universe and in the moderate-redshift
cluster CL 1358+62 suggest strongly that a large
fraction of early-type galaxies contain disc-like structures.

We have investigated how the lensing properties of early-type
galaxies would change if a thin disc component that contains about
5\,per cent of the mass is included: 

\begin{enumerate}
\item If early-type galaxies contain disc components, the statistical
lensing properties of early-type lensing galaxies are affected 
by less than about 5\,per cent. Only if the majority of lensing
galaxies are of S0/Sa or later types would the expected relative number
of quad to 2-image lens systems increase by a factor of 2 or more.
\item A disc component in early-type lensing galaxies affects
significantly the expected magnification ratios of highly magnified
quad systems.  If a disc component is present in the lens and
inclined at more than $\sim 70\,\deg$ to the line of sight, the
resulting image configurations and magnification ratios can resemble
that of systems like B1422+231. 
\item Time delays may be affected significantly by the presence of a
disc component in individual early-type lens galaxies. Bulge-only lens
models used to fit bulge-plus-disc lenses fit lensed image
positions well, but would yield a value for $H_0$ that is systematically
low by about 25\,per cent.

\end{enumerate}

The presence of
inclined exponential discs in early-type lensing galaxies affect the
lensing properties of quad lens systems significantly. 
Assuming that the majority of all early-type lens galaxies contain
discs, then both the time delays and magnification ratios are expected
to be affected by disc components in $10-30$\,per cent of all quad lens
systems. One or two of the quad lenses identified by
CLASS is expected to be affected (consistent with the properties of
B1422+231). 
Bulge-plus-disc models provide an alternative
explanation to the observed magnification ratios and lead to
systematically longer time-delays. In order to obtain more
stringent constraints on the presence of CDM substructure
in the lens and on the effect of discs on $H_0$ measurements, detailed lens
modelling of known lens systems will be necessary. Fitting models that 
include exponential discs to
real lens systems is much more computationally intensive than the 
calculations performed here, but may be possible. 

We have discussed the first order
effect of discs in early-type galaxies on observed lensing
properties. Once a large number of lens
systems is known, it might also be possible to use lensing to
constrain the masses and parameters of those systems. 
Determining the abundance and properties of discs in
ellipticals in more detail, and at different redshifts, using
gravitational lensing would shed new light on the details of the
formation of discs in early-type galaxies.

\section*{acknowledgements}
OM acknowledges financial support from a European Community Marie
Curie Fellowship. We thank Konrad Kuijken and the referee for helpful comments on the manuscript.



\appendix

\section{Galaxies in CL 1358+62}
All the galaxies for which \nocite{kelson2000}{Kelson} {et~al.} (2000) provides surface
brightness profile information are employed in our calculation of
certain statistical properties, like the high magnification cross
sections in \S\ref{cross}. For completeness, we present here the
surface brightness profile parameters for all the galaxies.

\begin{table*}
\begin{tabular}{l|r|r|r|r|r||r|r|r|r|r|r|r|r|r}\hline
 & & \multicolumn{1}{l|}{$\longleftarrow$} & \multicolumn{2}{c|}{Bulge only} & $\longrightarrow$ & \multicolumn{1}{l|}{$\longleftarrow$} & \multicolumn{7}{c|}{Bulge and disc} & $\longrightarrow$\\ \hline 
No. & Type & $<I_{\mathrm{h}}>$ & $r_{\mathrm{h}}$ & $\Sigma_{\mathrm{b}}$ & $R_{\mathrm{e}}$ & $<I_{\mathrm{h}}>$ & $r_{\mathrm{h}}$ & $\Sigma_{\mathrm{b}}$ & $<I_{\mathrm{hd}}>$ & $r_{\mathrm{d}}$ & $\Sigma_{\mathrm{d}}$ & $R_{\mathrm{e}}$ & $f_{\mathrm{e}}$ & $f_{\mathrm{t}}$\\ \hline
212 & E & 20.780 & 0.683 & 0.488 & 1.220 & 21.170 & 0.781 & 0.340 & 18.030 & 0.031 & 6.138 & 1.160 & 0.808 & 0.960\\
242 & E & 21.920 & 1.529 & 0.171 & 1.360 & 22.420 & 1.825 & 0.108 & 20.720 & 0.156 & 0.515 & 1.270 & 0.806 & 0.950\\
256 & E & 20.880 & 1.380 & 0.445 & 2.330 & 21.200 & 1.551 & 0.331 & 18.000 & 0.053 & 6.310 & 2.260 & 0.836 & 0.967\\
303 & E & 20.590 & 0.638 & 0.581 & 1.270 & 21.110 & 0.774 & 0.360 & 18.890 & 0.054 & 2.780 & 1.210 & 0.787 & 0.947\\
360 & E & 20.220 & 0.341 & 0.817 & 2.176 & 21.660 & 0.594 & 0.217 & 17.620 & 0.030 & 8.954 & 1.000 & 0.565 & 0.867\\
375 & E & 22.860 & 4.979 & 0.072 & 2.650 & 23.010 & 5.267 & 0.063 & 19.220 & 0.096 & 2.051 & 2.870 & 0.800 & 0.984\\
409 & E & 21.270 & 0.498 & 0.310 & 2.176 & 21.490 & 0.542 & 0.254 & 22.600 & 0.132 & 0.091 & 0.660 & 0.959 & 0.969\\
412 & E & 21.390 & 0.767 & 0.278 & 2.176 & 22.320 & 1.091 & 0.118 & 17.660 & 0.033 & 8.630 & 0.870 & 0.502 & 0.908\\
531 & E & 21.260 & 1.549 & 0.313 & 2.090 & 21.710 & 1.830 & 0.207 & 19.300 & 0.108 & 1.905 & 1.980 & 0.772 & 0.954\\
534 & E & 21.210 & 0.620 & 0.328 & 0.860 & 22.170 & 0.878 & 0.136 & 16.680 & 0.019 & 21.281 & 0.770 & 0.489 & 0.901\\
536 & E & 21.180 & 1.266 & 0.337 & 1.790 & 21.520 & 1.433 & 0.247 & 18.570 & 0.057 & 3.733 & 1.720 & 0.807 & 0.965\\
233 & E/S0 & 20.400 & 0.784 & 0.692 & 1.740 & 20.940 & 0.952 & 0.421 & 18.740 & 0.067 & 3.192 & 1.640 & 0.796 & 0.947\\
269 & E/S0 & 20.390 & 0.962 & 0.698 & 2.150 & 20.620 & 1.045 & 0.565 & 18.710 & 0.055 & 3.281 & 2.100 & 0.905 & 0.976\\
309 & E/S0 & 20.020 & 0.502 & 0.982 & 1.370 & 20.210 & 0.540 & 0.824 & 18.910 & 0.036 & 2.729 & 1.000 & 0.913 & 0.979\\
353 & E/S0 & 21.050 & 1.120 & 0.380 & 1.710 & 21.240 & 1.205 & 0.319 & 19.970 & 0.079 & 1.028 & 1.680 & 0.902 & 0.979\\
381 & E/S0 & 20.340 & 0.512 & 0.731 & 2.176 & 20.500 & 0.550 & 0.631 & 20.020 & 0.052 & 0.982 & 1.000 & 0.924 & 0.979\\
493 & E/S0 & 19.980 & 0.193 & 1.019 & 0.540 & 21.660 & 0.374 & 0.217 & 18.740 & 0.035 & 3.192 & 0.480 & 0.549 & 0.838\\
095 & S0 & 20.940 & 0.704 & 0.421 & 1.150 & 21.840 & 0.986 & 0.184 & 17.750 & 0.038 & 7.943 & 1.040 & 0.583 & 0.912\\
135 & S0 & 20.450 & 0.418 & 0.661 & 0.900 & 20.870 & 0.490 & 0.449 & 18.010 & 0.023 & 6.252 & 0.870 & 0.817 & 0.956\\
182 & S0 & 20.460 & 0.355 & 0.655 & 0.760 & 20.770 & 0.392 & 0.492 & 21.910 & 0.130 & 0.172 & 0.740 & 0.941 & 0.945\\
211 & S0 & 20.120 & 0.448 & 0.895 & 1.160 & 20.410 & 0.458 & 0.685 & 21.680 & 0.256 & 0.213 & 1.120 & 0.921 & 0.873\\
215 & S0 & 20.660 & 0.457 & 0.545 & 2.176 & 21.770 & 0.700 & 0.196 & 16.950 & 0.022 & 6.596 & 0.790 & 0.524 & 0.889\\
236 & S0 & 21.100 & 0.776 & 0.363 & 1.150 & 21.630 & 0.944 & 0.223 & 17.870 & 0.033 & 7.112 & 1.080 & 0.711 & 0.946\\
298 & S0 & 19.690 & 0.554 & 1.330 & 1.810 & 20.840 & 0.835 & 0.461 & 19.530 & 0.142 & 1.542 & 1.630 & 0.735 & 0.873\\
300 & S0 & 19.840 & 0.453 & 1.159 & 2.176 & 20.530 & 0.584 & 0.614 & 18.660 & 0.055 & 3.436 & 1.280 & 0.789 & 0.932\\
343 & S0 & 20.540 & 0.404 & 0.608 & 0.830 & 21.100 & 0.492 & 0.363 & 20.930 & 0.098 & 0.425 & 0.780 & 0.879 & 0.935\\
359 & S0 & 20.260 & 0.574 & 0.787 & 1.380 & 21.050 & 0.769 & 0.380 & 19.050 & 0.072 & 2.399 & 1.270 & 0.748 & 0.923\\
408 & S0 & 19.820 & 0.382 & 1.180 & 1.160 & 20.660 & 0.523 & 0.545 & 18.860 & 0.057 & 2.858 & 1.080 & 0.755 & 0.915\\
410 & S0 & 20.670 & 0.488 & 0.540 & 0.930 & 21.130 & 0.575 & 0.353 & 20.840 & 0.093 & 0.461 & 0.880 & 0.886 & 0.951\\
463 & S0 & 20.430 & 0.639 & 0.673 & 2.176 & 20.740 & 0.714 & 0.506 & 21.180 & 0.136 & 0.337 & 1.350 & 0.926 & 0.965\\
481 & S0 & 20.250 & 0.260 & 0.794 & 2.176 & 20.760 & 0.157 & 0.497 & 20.650 & 0.166 & 0.550 & 1.000 & 0.521 & 0.351\\
110 & S0/a & 19.450 & 0.350 & 1.660 & 1.300 & 18.360 & 0.102 & 4.529 & 20.120 & 0.248 & 0.895 & 1.000 & 0.722 & 0.363\\
129 & S0/a & 19.930 & 0.489 & 1.067 & 1.400 & 20.950 & 0.726 & 0.417 & 16.680 & 0.028 & 21.281 & 1.300 & 0.647 & 0.899\\
142 & S0/a & 21.400 & 0.834 & 0.275 & 1.030 & 20.470 & 0.318 & 0.649 & 22.440 & 0.648 & 0.106 & 0.870 & 0.934 & 0.496\\
164 & S0/a & 22.620 & 2.251 & 0.090 & 1.250 & 22.920 & 2.541 & 0.068 & 17.770 & 0.036 & 7.798 & 1.230 & 0.616 & 0.966\\
292 & S0/a & 20.320 & 0.486 & 0.745 & 1.130 & 20.390 & 0.297 & 0.698 & 21.100 & 0.332 & 0.363 & 1.000 & 0.803 & 0.506\\
335 & S0/a & 20.860 & 0.539 & 0.453 & 0.920 & 20.900 & 0.541 & 0.437 & 20.610 & 0.035 & 0.570 & 0.910 & 0.963 & 0.992\\
366 & S0/a & 20.790 & 0.630 & 0.483 & 1.120 & 21.160 & 0.718 & 0.344 & 20.460 & 0.083 & 0.655 & 1.070 & 0.882 & 0.963\\
369 & S0/a & 21.860 & 1.036 & 0.180 & 0.960 & 21.980 & 1.085 & 0.161 & 20.280 & 0.047 & 0.773 & 0.950 & 0.903 & 0.987\\
397 & S0/a & 20.500 & 0.409 & 0.631 & 0.860 & 18.960 & 0.111 & 2.606 & 21.450 & 0.320 & 0.263 & 1.000 & 0.841 & 0.443\\
423 & S0/a & 21.610 & 1.021 & 0.227 & 1.110 & 22.460 & 1.397 & 0.104 & 17.950 & 0.043 & 6.607 & 1.000 & 0.515 & 0.918\\
440 & S0/a & 22.450 & 1.108 & 0.105 & 0.690 & 22.980 & 1.328 & 0.064 & 18.800 & 0.036 & 3.020 & 0.640 & 0.572 & 0.950\\
454 & S0/a & 20.420 & 0.595 & 0.679 & 1.310 & 22.270 & 0.990 & 0.124 & 20.610 & 0.273 & 0.570 & 1.100 & 0.646 & 0.655\\
523 & S0/a & 21.640 & 1.044 & 0.221 & 1.110 & 21.960 & 1.171 & 0.164 & 21.080 & 0.117 & 0.370 & 1.070 & 0.882 & 0.968\\
209 & Sa & 20.650 & 0.625 & 0.550 & 1.200 & 22.060 & 0.960 & 0.150 & 20.930 & 0.253 & 0.425 & 1.040 & 0.755 & 0.772\\
328 & Sa & 20.720 & 0.666 & 0.515 & 1.230 & 21.530 & 0.862 & 0.244 & 21.300 & 0.232 & 0.302 & 1.130 & 0.862 & 0.881\\
356 & Sa & 20.200 & 0.604 & 0.832 & 1.500 & 19.780 & 0.392 & 1.225 & 22.070 & 0.533 & 0.149 & 1.420 & 0.939 & 0.749\\
368 & Sab & 21.100 & 0.542 & 0.363 & 2.176 & 22.390 & 0.627 & 0.111 & 21.560 & 0.305 & 0.238 & 1.000 & 0.713 & 0.568\\
371 & Sa & 21.040 & 0.956 & 0.384 & 1.470 & 20.640 & 0.468 & 0.555 & 21.920 & 0.693 & 0.171 & 1.260 & 0.890 & 0.497\\
372 & Sa & 21.920 & 1.435 & 0.171 & 1.280 & 22.720 & 1.905 & 0.082 & 21.990 & 0.351 & 0.160 & 1.150 & 0.877 & 0.909\\
465 & Sa & 21.010 & 0.609 & 0.394 & 0.950 & 21.220 & 0.660 & 0.325 & 21.100 & 0.079 & 0.363 & 1.000 & 0.925 & 0.977\\
549 & Sab & 21.920 & 0.866 & 0.171 & 0.770 & 19.900 & 0.142 & 1.096 & 22.500 & 0.609 & 0.100 & 0.560 & 0.938 & 0.284\\
234 & Sb & 21.860 & 0.996 & 0.180 & 0.920 & 17.800 & 0.035 & 7.586 & 22.030 & 0.621 & 0.154 & 0.520 & 0.903 & 0.094\\
\hline
\end{tabular}

\caption{Galaxies in CL 1358+62. The values for $<I_{\mathrm{b}}>$,
$r_{\mathrm{b}}$ and $<I_{\mathrm{hd}}>$ are taken
directly from Kelson et al. (2000). The disc scale length $r_{\mathrm{d}}$ is related to the half-light radius as given in Kelson et al. by $r_{\mathrm{d}}=1.688r_{\mathrm{hd}}$. In a cosmology with
$\Omega_{\Lambda}=0.7$, $\Omega_{\mathrm{M}}=0.3$ and
$H_0=50\,\mathrm{km}\,\mathrm{s}^{-1}\,\mathrm{Mpc}^{-1}$, $1\arcseca$
corresponds to $6.6\,\mathrm{kpc}$ at a redshift of $z=0.33$.}
\label{table.all}
\end{table*}

\end{document}